\documentclass[1p, final]{elsarticle}

\usepackage{subfig}
\usepackage{amsmath,bm}
\usepackage{color}
\usepackage{enumerate}
\usepackage{multirow}
\usepackage{amssymb}
\usepackage{epstopdf}
\usepackage{epsfig}
\usepackage[table,xcdraw]{xcolor}
\usepackage{caption}
\usepackage[linesnumbered,ruled,vlined]{algorithm2e} 

\usepackage{tikz}
\usepackage{graphicx}
\usetikzlibrary{positioning}

\usepackage{bigstrut}
\usepackage{lineno,hyperref}
\modulolinenumbers[5]

\journal{ }

\begin{document}

\begin{frontmatter}
	
\title{A Multigrid Graph U-Net Framework for Simulating Multiphase Flow in Heterogeneous Porous Media}

\author{Jiamin Jiang}
\ead{jiaminjiang@ustc.edu.cn}

\address{University of Science and Technology of China and Suzhou Institute for Advanced Research, Suzhou, China}

\author{Jingrun Chen}
\ead{jingrunchen@ustc.edu.cn}

\author{Zhouwang Yang}
\ead{yangzw@ustc.edu.cn}

\address{School of Mathematical Sciences and Suzhou Institute for Advanced Research, University of Science and Technology of China, China}

\begin{abstract}
	
Numerical simulation of multi-phase fluid dynamics in porous media is critical to a variety of geoscience applications. Data-driven surrogate models using Convolutional Neural Networks (CNNs) have shown promise but are constrained to regular Cartesian grids and struggle with unstructured meshes necessary for accurately modeling complex geological features in subsurface simulations.

To tackle this difficulty, we build surrogate models based on Graph Neural Networks (GNNs) to approximate space-time solutions of multi-phase flow and transport processes. Particularly, a novel Graph U-Net framework, referred to as AMG-GU, is developed to enable hierarchical graph learning for the parabolic pressure component of the coupled partial differential equation (PDE) system. Drawing inspiration from aggregation-type Algebraic Multigrid (AMG), we propose a graph coarsening strategy adapted to heterogeneous PDE coefficients, achieving an effective graph pooling operation.

Results of three-dimensional heterogeneous test cases demonstrate that the multi-level surrogates predict pressure and saturation dynamics with high accuracy, significantly outperforming the single-level baseline. Our Graph U-Net model exhibits great generalization capability to unseen model configurations.

\end{abstract}

\end{frontmatter}


\section{Introduction}


Multi-phase fluid dynamics in porous media is vital to a wide range of geoscience applications, including hydrocarbon recovery, aquifer management, and geological $\mathrm{CO_2}$ sequestration. Numerical simulations have become increasingly important for understanding, quantifying, and controlling subsurface flow processes. Detailed geological models with heterogeneous properties commonly serve as inputs to numerical solvers. Forecasting the evolution of fluid dynamics involves solving the parameterized partial differential equations (PDEs) governing multi-phase flow and transport. These PDEs are often highly nonlinear and exhibit a complex interplay of parabolic and hyperbolic characteristics. Consequently, the development of computationally efficient and accurate simulation techniques is essential for applications in Earth's subsurface.


The emergence of deep learning has significantly influenced various scientific fields, notably computer vision and natural language processing. Convolutional Neural Networks (CNNs) have become foundational for processing grid-like data structures such as images (LeCun et al. 1998; Krizhevsky et al. 2012). One of the most impactful variants of CNNs is the U-Net architecture (Ronneberger et al. 2015), which has proven exceptionally effective in pixel-wise prediction tasks. U-Nets operate by progressively building hierarchical representations of the input data, employing pooling layers to enlarge the receptive fields of neurons (Scherer et al. 2010). These pooling layers enable the network to capture multi-scale patterns from the data. This not only reduces computational complexity—facilitating the construction of wider and deeper networks—but also enhances generalization and performance by focusing on the most salient features.

In recent years, a substantial body of research has successfully utilized CNN architectures to develop data-driven surrogate models for approximating solutions of PDEs, particularly in the field of fluid dynamics (Guo et al. 2016; Bar-Sinai et al. 2019; Mo et al. 2019; Bhatnagar et al. 2019; Sekar et al. 2019; Li et al. 2020; Ribeiro et al. 2020; Thuerey et al. 2020; Santos et al. 2020; Liu et al. 2020; Lu et al. 2021; Jiang et al. 2021). Compared to high-fidelity numerical solvers, a learned simulator can offer much faster predictions, especially for high-dimensional physics problems.

In the realm of subsurface flow and transport modeling, a number of studies have applied U-Nets and snapshots of simulation data to learn the nonlinear mappings from the input rock properties to the output states (pressure and saturation) on regular Cartesian meshes (Tang et al. 2020; Wen et al. 2021; Maldonado-Cruz and Pyrcz 2022; Yan et al. 2022; Zhang et al. 2022; Wen et al. 2022; Jiang and Durlofsky 2023). While CNNs are effective at approximating PDE solutions, they are restricted to a specific discretization of the physical domain on which they are trained. Due to the inherent limitations of standard convolution operations, it remains challenging for CNNs to handle irregular and unstructured simulation meshes. Given that subsurface simulation models often require corner-point and unstructured meshes with skewed and degenerate geometries to accurately characterize complex geological features and heterogeneity, these complexities limit the applicability of CNN-based models in subsurface problems.


Graph Neural Networks (GNNs) have shown tremendous potential for learning mesh-based simulations of time-dependent PDE systems (Li et al. 2020; Iakovlev et al. 2020; Belbute-Peres et al. 2020; Chen et al. 2021; Brandstetter et al. 2022; Peng et al. 2022; Lam et al. 2023; Franco et al. 2023; Wang et al. 2023). Particularly, Pfaff et al. (2020) developed a GNN named MeshGraphNet (MGN), using an encoder-processor-decoder structure to accurately simulate a broad range of physical phenomena, including fluid dynamics and structural mechanics. Jiang (2024) proposed a GNN framework based on graph transformer to efficiently simulate multi-phase dynamics in fractured media, demonstrating good generalizability for varying fracture geometries. Ju et al. (2024) integrated a graph-based convolutional Long-Short-Term Memory model with MGN to forecast $\mathrm{CO_2}$ plume migration during geological storage in faulted reservoirs. The model can reduce temporal error accumulation and improve forecast accuracy. GNNs are naturally capable of operating on unstructured meshes with complex geometries, as opposed to CNNs. A simulation mesh can be abstracted as a graph composed of nodes and edges linking each node pair. The core mechanism of GNNs involves propagating and aggregating local information into node representations, through multiple message-passing layers (Kipf and Welling 2016; Gilmer et al. 2017).


Building on the success of U-Nets in image-based applications, researchers have adapted these architectures to graph data, leading to the development of Graph U-Nets (Gao and Ji 2021). GNNs pose unique challenges for implementing pooling and upsampling operations due to the absence of spatial locality and the irregularity of node connections. To overcome this, graph pooling modules have been introduced to generate hierarchical representations of graphs (Ying et al. 2018; Lee et al. 2019). These modules enable GNNs to capture multi-scale dependencies, enhancing the model ability to learn complex patterns. Recently, Graph U-Nets have shown potential in physics-based simulations. Specifically, Fortunato et al. (2022) enhanced MeshGraphNets with a two-scale framework, to handle high-resolution simulations by leveraging multi-scale message passing. Lino et al. (2022) presented a multi-scale rotation-equivariant GNN for unsteady Eulerian fluid dynamics, enhancing the model ability to generalize across fluid systems with varying Reynolds numbers. They performed graph coarsening using a VoxelGrid method, which assigns fine nodes to a regular grid. Cao et al. (2022) developed a bi-stride multi-scale GNN for physics simulations on large-scale meshes, introducing a pooling strategy that enhances connectivity preservation and computational efficiency. Deshpande et al. (2024) devised a Graph U-Net architecture for surrogate modeling in nonlinear finite element simulations.


In the present work, we construct GNN-based surrogate models to approximate space-time solutions of multi-phase flow and transport processes in porous media. Since the coupled PDE system has mixed parabolic and hyperbolic characters, we apply specific network architectures suitable for the different characters. To achieve effective hierarchical graph learning of the parabolic pressure component, we develop a novel Graph U-Net framework that can capture multi-scale features and reduce computational complexity. Our graph learning task targets a model with spatially varying PDE coefficients, which, to the best of our knowledge, has not been addressed in prior works. Inspired by aggregation-type Algebraic Multigrid (Vanek et al. 1996; Muresan and Notay 2008), we propose a graph coarsening strategy adapted to heterogeneous information, achieving an effective graph pooling operation.

We evaluate the model performance using three-dimensional heterogeneous test cases. The results show that our multi-level surrogates predict pressure and saturation dynamics with high accuracy, significantly outperforming the single-level baseline. The AMG-inspired Graph U-Net effectively captures both local details and long-range patterns that are essential for the pressure dynamics. The multi-level representation provides the surrogates with better generalization to unseen model configurations.


\section{Problem definition}

\subsection{Overview}

We first introduce a generic form of time-dependent nonlinear parameterized partial differential equations (PDEs)
\begin{align}
	& \frac{\partial \bm{u}(t, \bm{x})}{\partial t} = \mathcal{F} \left( \bm{x}, \bm{u}, \nabla \bm{u}, \nabla^2 \bm{u}; \boldsymbol{\varphi} \right), \qquad (t, \bm{x}) \in [0, T] \times \Omega  \\
	& \bm{u}(t = 0, \bm{x}) = \bm{u}^0(\bm{x}), \qquad \bm{x} \in \Omega  \\
	& \bm{u}(t, \bm{x}) = \bar{\bm{u}}(t, \bm{x}), \qquad \quad \ (t, \bm{x}) \in [0, T] \times \partial \Omega_{D}  \\
	& \bar{\bm{f}}\big(\nabla \bm{u}, \bm{n}(\bm{x})\big) = 0. \qquad \quad (t, \bm{x}) \in [0, T] \times \partial \Omega_{N}
\end{align}
which are widely applicable to modelling of fluid dynamics. $\bm{u}(t, \bm{x})$ represents the state variables, $\bm{x} \in \Omega$ are the spatial coordinates, $\mathcal{F}$ is a nonlinear function of various differential operators, and $\boldsymbol{\varphi}$ are the PDE coefficients. The PDE system is subjected to Dirichlet and Neumann boundary conditions, which are defined on $\partial \Omega_{D}$ and $\partial \Omega_{N}$, respectively, and $\bm{n}(\bm{x})$ is the unit normal vector to $\partial \Omega_{N}$.

One popular way of solving such PDE model is to apply finite volume methods, which partition the simulation domain $\Omega$ into an unstructured mesh $\left\{ \Omega_i \right\}_{i=1}^{n}$ consisting of $n$ cells. At time $t^m$, the discrete state vector $\bm{U}^{m} = \left\{ \bm{u}_i^m \right\}_{i=1}^{n}$ can thus be defined by the state value $\bm{u}_i^m$ at each cell center. With a fully-implicit scheme (first-order backward Euler) for time-stepping, the discretized nonlinear system can be written as
\begin{equation} 
	\bm{R}\left ( \bm{U}^{m+1}; \boldsymbol{\varphi} \right ) = 0 
\end{equation}
where $\bm{R}$ represents the residual vector. The nonlinear system is often solved using Newton's method. For each timestep, with the last solution $\bm{U}^{m}$, and a timestep size $\Delta t$, the new state $\bm{U}^{m+1}$ will be acquired.

We denote a high-fidelity simulator as $\mathbb{H}$ that maps the current state of mesh cells to the next timestep state. A rollout trajectory of states $\left ( \bm{U}^{0}, \bm{U}^{1}, ..., \bm{U}^{n_t} \right )$ can be computed iteratively by applying $\bm{U}^{m+1} = \mathbb{H} \left ( \bm{U}^{m} \right )$ over $n_t$ timesteps.

Our data-driven learning task is to replace the computationally expensive high-fidelity simulator with surrogate simulators that predict the next state 
\begin{equation} 
	\bm{U}^{m+1} \approx \widehat{\bm{U}}^{m+1} = \mathbb{N} \left ( \bm{U}^{m}; \Theta \right )
\end{equation}
where $\mathbb{N}$ is a one-step prediction model based on GNNs, whose parameters $\Theta$ will be optimized for certain end-to-end training objectives. $\widehat{\bm{U}}^{m+1}$ indicates the predicted state from the surrogate model. Given the initial state $\bm{U}^{0}$, the rollout trajectory $\left ( \bm{U}^{0}, \widehat{\bm{U}}^{1}, ..., \widehat{\bm{U}}^{n_t} \right )$ can be rapidly produced at inference time through $\mathbb{N} \left ( \cdot ; \Theta \right )$ in an autoregressive way.

\subsection{Multi-phase flow in porous media}

In this work, we focus on the compressible and immiscible flow and transport problem in porous media with $n_p$ number of phases. The mass balance equation for phase $l \in \left \{ 1,...,n_p \right \}$ can be expressed as
\begin{equation} 
	\label{eq:mass_con}
	\frac{\partial }{\partial t } \left ( \phi \rho_{l} s_{l} \right ) + \nabla \cdot \left (\rho_{l} \bm{v}_{l} \right ) - \rho_{l} q_{l} = 0,
\end{equation}
where $q_{l}$ is source or sink term, describing injection or extraction of fluids from reservoir, and $s_{l}$ is fluid phase saturation, which is constrained by $\sum_{l} s_{l} = 1$. The Darcy phase velocity is given as
\begin{equation} 
	\label{eq:phase_vel}
	\bm{v}_l = -K \lambda_l \left ( \nabla p_l - \rho_l g \nabla z \right ).
\end{equation}
where rock permeability $K$ can be viewed as time independent diffusion coefficient that varies in space. $p_l$ is phase pressure, $g$ is the gravitational acceleration constant, and $z$ is depth (assuming positive downward). $\lambda_l = K_{rl}/\mu_l$ is phase mobility, where $K_{rl}$ and $\mu_l$ are relative permeability and fluid viscosity, respectively.

The above system is called the fully coupled formulation. The independent primary variables are one phase pressure and $(n_p-1)$ phase saturations. Rock porosity $\phi$ and phase density $\rho_{l}$ are nonlinearly dependent on the pressure, and $k_{rl}$ is a nonlinear function of the saturations. The fully-implicit finite-volume discretization for the fully coupled model are summarized in Appendix A.

For a system that only involves two fluid (nonwetting and wetting) phases, Eq.~(\ref{eq:mass_con}) can be simplified to
\begin{align}
	& \frac{\partial}{\partial t } \left ( \phi \rho_{nw} s_{nw} \right ) + \nabla \cdot \left ( \rho_{nw} \bm{v}_{nw} \right ) - \rho_{nw} q_{nw} = 0, \\
	& \frac{\partial}{\partial t } \left ( \phi \rho_{w} s_{w} \right ) + \nabla \cdot \left ( \rho_{w} \bm{v}_{w} \right ) - \rho_{w} q_{w} = 0, 
\end{align}
with the saturation constraint as $s_{nw} + s_w = 1$. The corresponding capillary pressure is defined as the difference between the phase pressures
\begin{equation}
	p_{ca}(s_w) = p_{nw} - p_{w}.
\end{equation}

\subsubsection{Fractional flow formulation}

It is usually advantageous to reorganize Eq.~(\ref{eq:mass_con}) into one parabolic equation for the pressure and a system of hyperbolic equations for saturations (Chen et al. 2006). Specialized solvers suitable to the corresponding characters of the equations can be employed to optimize computational performance.

We will now introduce the so-called fractional flow formulation (Chavent and Jaffré 1986; Chen and Ewing 1997). By summing the phase mass balance equations (\ref{eq:mass_con}), a pressure equation can be derived as
\begin{equation}
	\frac{\partial \phi}{\partial t} + \nabla \cdot \bm{v}_\mathrm{\top} + \sum_{l} \frac{1}{\rho_l} \left( \phi s_l \frac{\partial \rho_l}{\partial t} + \bm{v}_l \cdot \nabla \rho_l \right) - \sum_{l} q_l = 0,
\end{equation}
where $\bm{v}_\mathrm{\top} = \sum_{l} \bm{v}_l$ is the total velocity.

For the model with two phases, $\bm{v}_\mathrm{\top}$ can be expressed in terms of the nonwetting phase pressure as
\begin{equation}
	\bm{v}_\mathrm{\top} = -K\lambda_\mathrm{\top} \nabla p_{nw} + K\lambda_{w} \nabla p_{ca} + K\left( \lambda_{nw} \rho_{nw} + \lambda_{w} \rho_{w} \right) g \nabla z,
\end{equation}
where $\lambda_\mathrm{\top} = \lambda_{nw} + \lambda_{w}$ is the total mobility.

One of the phase mass balance equations can be selected as the saturation equation
\begin{equation}
	\frac{\partial}{\partial t } \left ( \phi \rho_{w} s_{w} \right ) + \nabla \cdot \left ( \rho_{w} \bm{v}_{w} \right ) - \rho_{w} q_{w} = 0,
\end{equation}
where the phase velocity is rewritten in terms of the total velocity
\begin{equation}
	\label{eq:fff_vw}	
	\bm{v}_{w} = \frac{\lambda_{w}}{\lambda_\mathrm{\top}} \bm{v}_\mathrm{\top} + K \frac{\lambda_{w} \lambda_{nw}}{\lambda_\mathrm{\top}} \Big ( \nabla p_{ca} + \left ( \rho_{w} - \rho_{nw} \right ) g \nabla z \Big ).
\end{equation}
Note that the saturation (transport) equation is an advection-dominated type, and is generally characterized by nonconvex and nonmonotonic flux functions in the presence of combined viscous, gravitational and capillary forces.

\subsubsection{Surrogate modelling}

The coupled multi-phase flow system has an intricate blend of parabolic and hyperbolic characters. Therefore in this work we individually design and train two GNN models that compute the solutions of pressure and saturation
\begin{equation}
	\label{eq:cm_PS}
	\left\{
	\begin{aligned}
		\widehat{\bm{p}}^{m+1} & = \mathbb{N}_p \left ( \bm{p}^{m}, \bm{s}^{m}; \Theta_p \right ) , \\ 
		\widehat{\bm{s}}^{m+1} & = \mathbb{N}_s \left ( \bm{p}^{m}, \bm{s}^{m}; \Theta_s \right ) .
	\end{aligned}
	\right.
\end{equation}
where $\mathbb{N}_p$ and $\mathbb{N}_s$ denote respectively the pressure and saturation models. At each timestep, both the models take the input from the previous timestep.

\section{Graph Neural Networks}

We rely on GNNs to construct data-driven surrogate simulators to approximate the PDE solutions. GNNs offer a flexible and efficient way to operate over graph-structured data, naturally fitting mesh-based simulations (Pfaff et al. 2020).

\begin{figure}[!htb]
\centering
\includegraphics[scale=0.42]{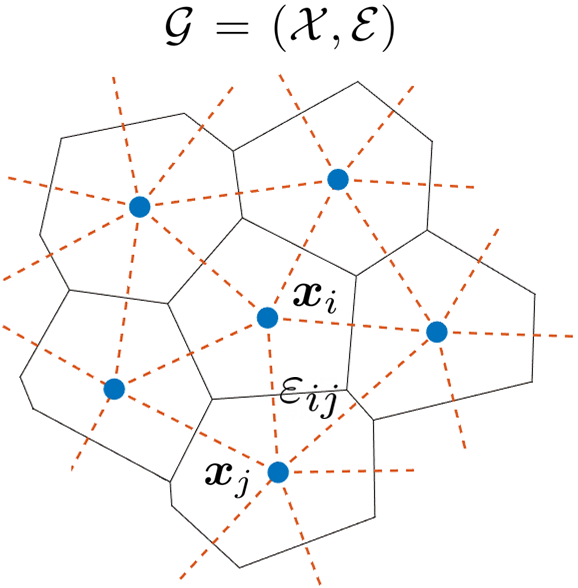}
\caption{Schematic of the graph representation of a typical finite-volume simulation mesh. Blue dots indicate nodes (cell centroids), and orange segments indicate edges (cell connections).}
\label{fig:graph_representation}
\end{figure}

A simulation mesh can be converted into a graph $\mathcal{G} = \left ( \mathcal{X}, \mathcal{E} \right )$ (\textbf{Fig.~\ref{fig:graph_representation}}) with nodes $\mathcal{X}$ (blue dots), undirected edges $\mathcal{E}$ (orange line segments), and an adjacency matrix $\bm{A}$ comprising edge connectivity. Here each node $i \in \mathcal{X}$ corresponds to the mesh cell $\Omega_i$. Let $\bm{x}_i$ be the cell centroid, and $\varepsilon_{ij}$ represents the edge for the connecting neighboring cells at $\bm{x}_i$ and $\bm{x}_j$. We denote $\mathcal{N}(i)$ as the set of adjacent nodes around node $i$. We further assign $\bm{h}_i$ and $\bm{e}_{ij}$ as the node and edge feature vectors respectively.

A GNN model consists of multiple generalized convolution layers, each aiming to propagate and aggregate local information across the one-hop neighborhood of each node (Battaglia et al. 2016; Kipf and Welling 2016). By stacking $\ell$ layers, the network can build node representations from the $\ell$-hop neighborhood. In a mesh-based fluid dynamics simulation, this allows each mesh node to incorporate information about the flow field from surrounding regions.

The core operation in GNNs is message passing, which updates the feature vectors of target nodes from the features of their neighbors. The general form of the message-passing scheme can be described by (Gilmer et al. 2017)
\begin{equation}
	{\bm{h}}'_i = \gamma \left ( \bm{h}_i, \underset{j\in\mathcal{N}(i)}{\bigoplus} \psi \left ( \bm{h}_i, \bm{h}_j, \bm{e}_{ij} \right ) \right ),
\end{equation}
where $\psi$ is a differentiable function responsible for constructing the message to be aggregated. One simple choice is given as
\begin{equation}
	\bm{\Pi}_{ij} = \psi \left ( \bm{h}_i, \bm{h}_j, \bm{e}_{ij} \right ) = \mathtt{MLP} \big ( [ \bm{h}_i \| \bm{h}_j \| \bm{e}_{ij} ] \big ).
\end{equation}
where $\Vert$ represents vector concatenation. The message depends on the features of both the source node $j$ and the target node $i$, as well as any edge features $\bm{e}_{ij}$. After the messages from each neighboring node are constructed, an aggregated information $\bm{\Pi}_{i}$ is created using a permutation invariant aggregation operation $\bigoplus$ (such as sum, mean, or max). The new feature vector ${\bm{h}}'_i$ is obtained by an update function $\gamma$ (such as MLP) involving a nonlinear transformation to the concatenation of the central features $\bm{h}_i$ and $\bm{\Pi}_{i}$. Each subsequent message-passing layer contains a separate set of network parameters, and operates on the output of the previous layer.

In this work, we will utilize a popular message-passing-based graph convolution operator as fundamental building blocks for Graph U-Net architectures.

\subsection{GATConv}

We consider the graph attentional (GAT) operator (Veličković et al. 2017), which adopts a self-attention process (Bahdanau et al. 2014) into graph learning. The attention mechanism allows a network to weigh the importance of different neighbors, enabling it to focus on more important information within the data.

In GAT, a shared linear transformation is first applied to every node. Then the self-attention using a shared attentional mechanism is performed, assigning an unnormalized coefficient for every node pair $(j, i)$ as
\begin{equation}
	\xi_{ij} = \text{LeakyReLU}\left(\mathbf{a}^{T}[\mathbf{W} \bm{h}_i \| \mathbf{W} \bm{h}_j]\right),
\end{equation}
which specifies the importance of node $j$'s features to node $i$. $\mathbf{W}$ is a learnable weight matrix. Here the attention mechanism is a single-layer feed-forward neural network, parametrised by a weight vector $\mathbf{a}$, and followed by the LeakyReLU nonlinearity with negative slope 0.2.

To incorporate multi-dimensional edge features, $\xi_{ij}$ may be computed as
\begin{equation}
	\xi_{ij} = \text{LeakyReLU}\left(\mathbf{a}^{T}[\mathbf{W}_1 \bm{h}_i \| \mathbf{W}_1 \bm{h}_j \| \mathbf{W}_2 \bm{e}_{ij}]\right),
\end{equation}

These coefficients are then normalised through softmax, in order to be comparable over different neighborhoods
\begin{equation}
\alpha_{ij} = \textrm{softmax}_j(\xi_{ij}) = \frac{\textrm{exp} \left ( \xi_{ij} \right )}{\sum_{k \in \mathcal{N}(i)\cup\left \{ i \right \}} \textrm{exp} \left ( \xi_{ik} \right ) },
\end{equation}
The attention scores are finally expressed as
\begin{equation}
	\alpha_{ij} = \frac{\textrm{exp} \Big ( \text{LeakyReLU}\left(\mathbf{a}^{T}[\mathbf{W}_1 \bm{h}_i \| \mathbf{W}_1 \bm{h}_j \| \mathbf{W}_2 \bm{e}_{ij}]\right) \Big )}{\sum_{k \in \mathcal{N}(i)\cup\left \{ i \right \}} \textrm{exp} \Big ( \text{LeakyReLU}\left(\mathbf{a}^{T}[\mathbf{W}_1 \bm{h}_i \| \mathbf{W}_1 \bm{h}_k \| \mathbf{W}_2 \bm{e}_{ik}]\right) \Big ) },
\end{equation}

After obtaining the attention scores, we can compute the nodal output (new node features) of a network layer as a weighted sum of the transformed features
\begin{equation}
	{\bm{h}}'_i = \sigma \left ( \alpha_{ii} \mathbf{W}_1 \bm{h}_i + \! \sum_{j\in\mathcal{N}(i)} \! \alpha_{ij} \mathbf{W}_1 \bm{h}_j \right ).
\end{equation}
where $\sigma$ denotes a nonlinear activation function, e.g., ReLU or Tanh.

\begin{figure}[!htb]
\centering
\includegraphics[scale=0.39]{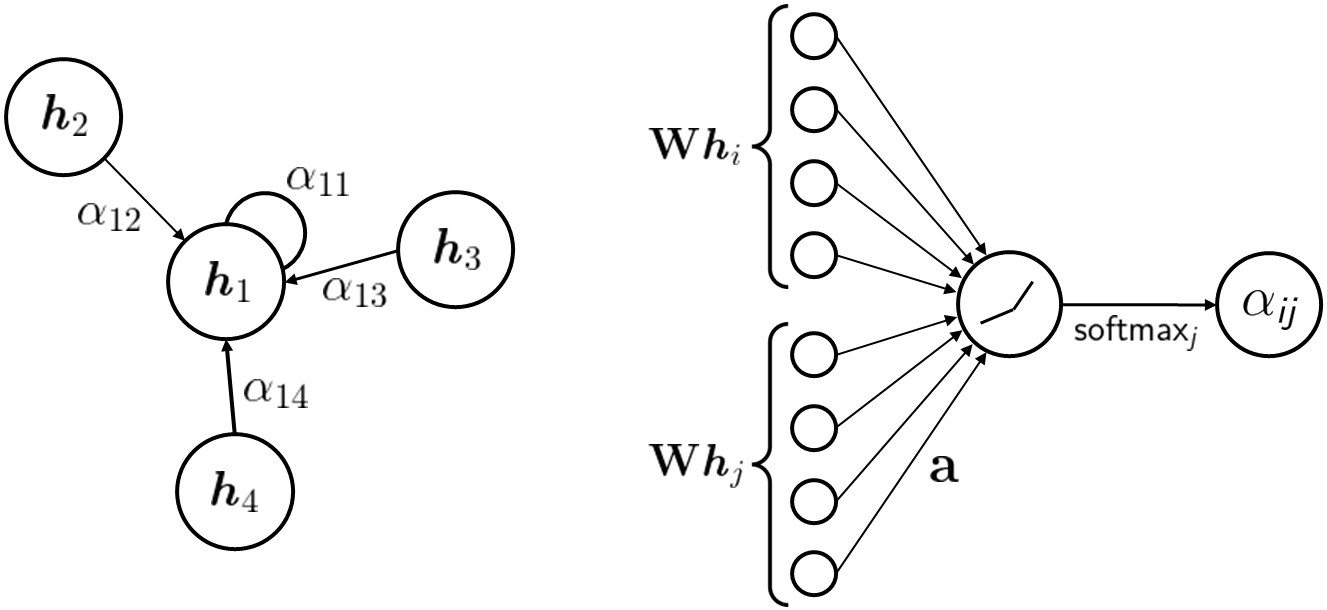}
\caption{Illustration of the GAT operator. Left: the importance of the neighboring nodes to node 1; Right: the attention scores from the attention mechanism.}
\label{fig:GAT_plot}
\end{figure}

An illustration of the GAT operator is shown in \textbf{Fig.~\ref{fig:GAT_plot}}.

\subsection{Model architectures}
\label{onelevel_gat}

\begin{figure}[!htb]
	\centering
	\includegraphics[scale=0.43]{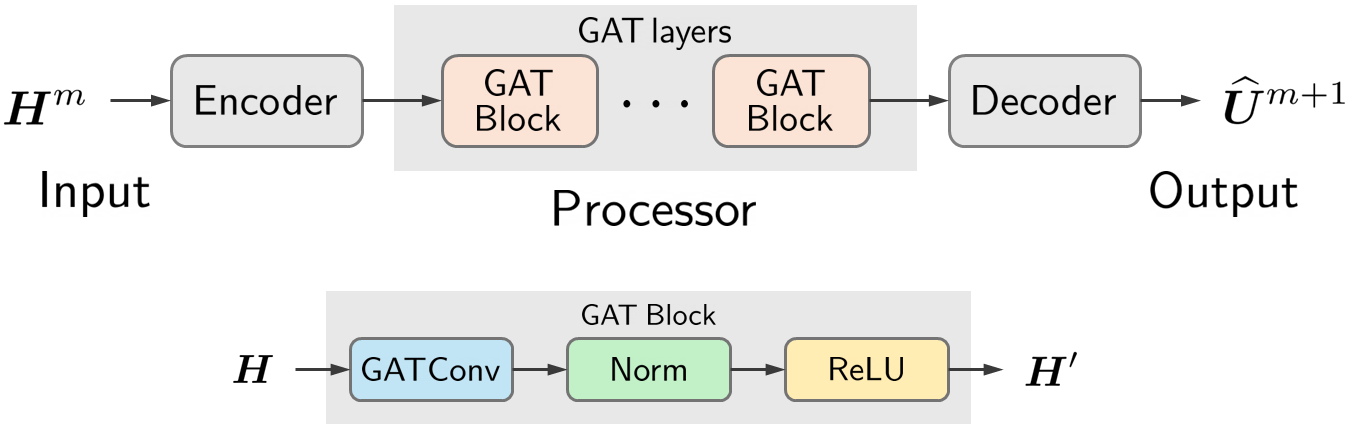}
	\caption{Schematic of a general GNN model architecture (single-level).}
	\label{fig:model_architecture}
\end{figure}

In this section, we present the details of the surrogate models that predict the next-step dynamic states of the coupled PDE system. Our GNN models have an Encoder-Processor-Decoder structure (Battaglia et al. 2018). Schematic of a general GNN model architecture (single-level) is plotted in \textbf{Fig.~\ref{fig:model_architecture}}.

For a mesh graph, the input features $\bm{h}_i$ of node $i$ at each timestep contain the dynamic variables (pressure and water saturation), permeability, and pore volume. A one-hot vector indicating node type (distinguishing reservoir, production, and injection nodes), along with the well index are also included. The initial nodal features are first encoded into latent vectors of size $n_h$. We denote the node feature matrix by $\bm{H} \in \mathbb{R}^{n \times n_h}$, where each row vector corresponds to the feature vector $\bm{h}_i$.

In the $\mathtt{GATConv}$ operator, the transmissibility $\Upsilon_{ij}$ of each connection (see Appendix A) and the vector of relative node positions $(\bm{x}_{ij} = \bm{x}_i - \bm{x}_j)$ are encoded as the multi-dimensional edge features. The relative positional information enables GNNs to possess spatial equivariance, which can be important for correctly capturing directional flow patterns induced by the transport problem (Iakovlev et al. 2020). All the input and target features are scaled individually to $[0, 1]$ by the min-max normalization.

The Decoder extracts the nodal field output of one target state (either $\widehat{\bm{p}}^{m+1} $ or $\widehat{\bm{s}}^{m+1} $) from the latent node features after the final processing layer. The Encoder and Decoder are two-layer MLPs with ReLU nonlinearities except for the output layer of the Decoder, after which we do not apply any nonlinearity.

The Processor of the pressure model $\mathbb{N}_p$ is constructed by stacking 9 identical GAT blocks, to obtain a sequence of updated latent features. Each block consists of one $\mathtt{GATConv}$ layer with the mean aggregation operation, the Layer Normalization $\mathtt{LayerNorm}$ (Ba et al. 2016), and the ReLU function. For the $\mathbb{N}_s$ model, we propose a combined architecture (2 $\mathtt{EdgeConv}$ followed by 4 GAT blocks with max aggregation), which is found to be quite effective at resolving the hyperbolic (saturation) solutions (Jiang 2024). The sizes of hidden units in $\mathbb{N}_p$ and $\mathbb{N}_s$ are $n_{hp} = 32$ and $n_{hs} = 128$, respectively.

\section{Graph U-Nets framework}


Graph U-Nets allow for the development of deep GNNs that can learn to operate on hierarchical representations of input graphs (Gao and Ji 2021). Similar to CNNs, pooling and unpooling operations for information transitions are essential building blocks in these architectures, enabling a model to capture multi-scale graph features and reduce computational complexity.


Algebraic multigrid (AMG) is one of the most efficient solvers for large-scale sparse linear systems arising from discretized PDEs (Brandt 1986; Ruge and Stüben 1987). AMG methods construct a hierarchy of coarser problems, relying on transfer operators to map between adjacent levels in this hierarchy. They execute through recursively applying a two-grid scheme, which combines smoothing iterations (e.g., Gauss-Seidel) with a coarse-grid correction. By solving a coarser problem with fewer variables, the coarse-grid correction resolves (smooth) low frequency errors that the fine-grid smoother (which primarily removes high frequency errors) overlooks.

\begin{figure}[!htb]
	\centering
	\includegraphics[scale=0.42]{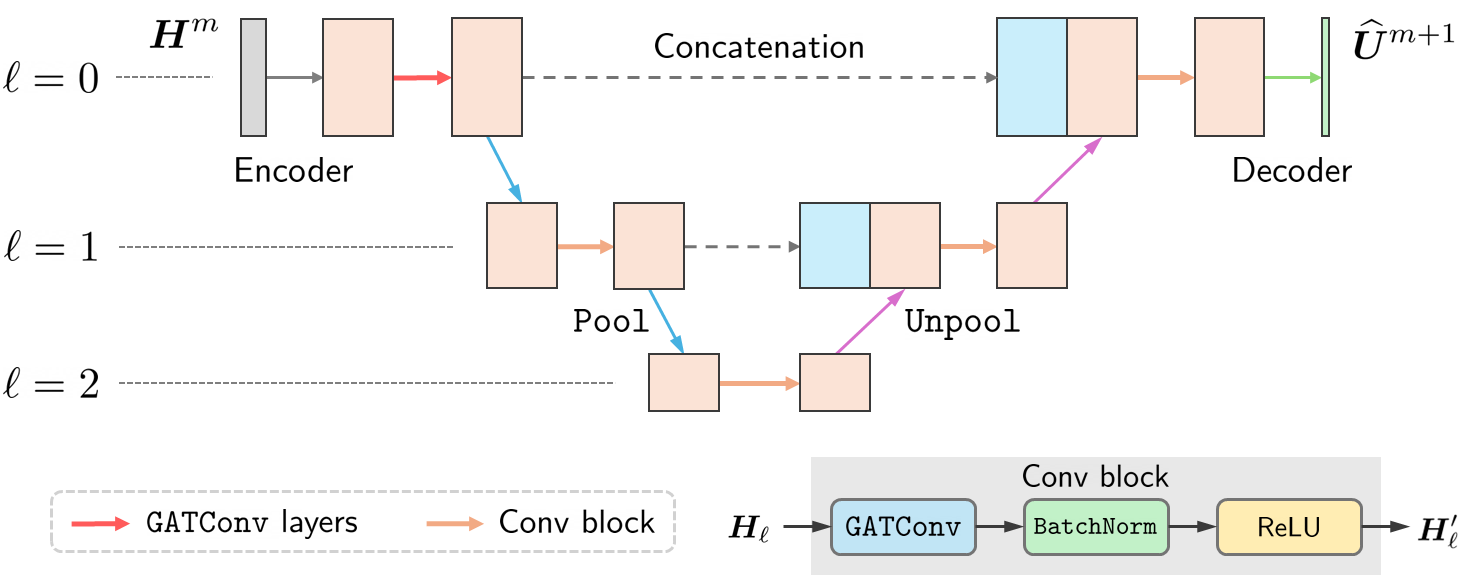}
	\caption{Schematic of a general Graph U-Net model architecture (the rectangles illustrate the dimensions of the node feature matrices, where the height represents the number of nodes, and the width represents the number of features).}
	\label{fig:GU_architecture}
\end{figure}

Illustration of a general Graph U-Net with the Encoder-Processor-Decoder structure is shown in \textbf{Fig.~\ref{fig:GU_architecture}}. As we can see, Graph U-Net closely resembles the standard multigrid V-cycle algorithm. At the finest (zeroth) level of Graph U-Net, the initial graph convolution layers are analogous to the pre-smoothing step of AMG.

In the downward (downsampling) path, there are several pooling blocks, each of which contains a $\mathtt{Pool}$ layer followed by a Conv block. The $\mathtt{Pool}$ layers create progressively coarser graphs to capture higher-order features (Zhang et al. 2018), while the Conv blocks aggregate information from the neighborhoods, until the coarsest level is reached. The upward (upsampling) path is constructed with an equal number of unpooling blocks as in the downsampling path. Each unpooling block is composed of an $\mathtt{Unpool}$ layer and a Conv block. The $\mathtt{Unpool}$ layer utilizes the information from the corresponding $\mathtt{Pool}$ layer and performs the inverse operation, restoring (interpolating) the graph into its higher resolution structure.

Like U-Nets on images, a skip connection (concatenation) is established at each level, to fuse features of the corresponding Conv blocks (Gao and Ji 2021). The skip-connections facilitate spatial information transmission for better model performance (Ronneberger et al. 2015). Finally, several convolution layers are stacked for final predictions, mirroring the post-smoother of AMG.

In this work, we develop a Graph U-Net architecture for $\mathbb{N}_p$, with pooling and unpooling operators built upon aggregation-type AMG, to achieve effective hierarchical graph learning of the pressure component in the multi-phase problem.

\subsection{Graph pooling}

\subsubsection{Review}

Inspired by the success of conventional pooling layers in CNNs, numerous recent works in the field of graph deep learning have introduced pooling operators as a means of forming coarse representations of input graphs. These pooling operators aim to identify the most important nodes and edges in the graph while preserving the underlying structural information (Grattarola et al. 2022; Liu et al. 2022).

Consider a graph $\mathcal{G}$ with $n$ nodes, each of which has $n_h$ features. The graph can also be represented by a node feature matrix $\bm{H} \in \mathbb{R}^{n \times n_h}$ and a weighted adjacency matrix $\bm{A} \in \mathbb{R}^{n \times n}$ ($\text{nodes} \ j \rightarrow i \ \text{are connected if} \ \bm{A}_{ij} \neq 0$). Each row vector in the feature matrix $\bm{H}$ corresponds to the feature vector $\bm{h}_i$ of node $i$ in the graph. A graph pooling operator can be defined as any function that maps a graph $\mathcal{G}$ to a (new) pooled graph $\mathcal{G}_c = \left ( \bm{H}_c, \bm{A}_c \right )$ with the generic goal of reducing the number of nodes from $n$ to $n_c < n$. The pooling ratio is defined as $(n_c/n)$, determining the degree of graph coarsening.


One family of graph pooling methods is node drop pooling, where a portion of nodes are discarded to form a coarsened graph (Lee et al. 2019; Gao and Ji 2021). In these methods, node features are projected to a scoring vector through a learnable transformation, and the scoring vector is then used to rank nodes for dropping. The coarsening operations of a node drop pooling can be generally formulated as
\begin{equation}
	\bm{y} = \mathtt{score} \left ( \bm{H}, \bm{A} \right ), \quad \mathrm{idx} = \mathtt{rank} \left ( \bm{y} \right ),
\end{equation}
\begin{equation}
	\bm{H}_{c} = \bm{H} \left ( \mathrm{idx}, : \right ) \in \mathbb{R}^{n_c \times n_h}, \quad \bm{A}_{c} = \bm{A} \left ( \mathrm{idx}, \mathrm{idx} \right ) \in \mathbb{R}^{n_c \times n_c}.
\end{equation}
where functions $\mathtt{score}$ and $\mathtt{rank}$ are designed for score generation and node selection, respectively. Based on the significance scores $\bm{y}$, the nodes with top-$k$ values are selected. The indices of the selected nodes are stored in $\mathrm{idx}$. Then the coarsened graph is constructed with the new node feature matrix $\bm{H}_{c}$ and adjacency matrix $\bm{A}_{c}$. Some prior research has revealed that the node-drop mechanism may suffer from the loss of node and edge information, potentially leading to performance degradation under various node-level and graph-level tasks (Bianchi et al. 2020; Grattarola et al. 2022). The unpooling operation for the family of node drop methods is commonly formulated as 
\begin{equation}
	\bm{H} = \mathtt{distribute} \, \big ( \mathbf{0}_{n \times n_h}, \bm{H}_{c}, \mathrm{idx} \big ) \in \mathbb{R}^{n \times n_h}
\end{equation}
where $\mathbf{0}_{n \times n_h}$ is the initial empty feature matrix for the fine graph. The row vectors of $\bm{H}$ with indices in $\mathrm{idx}$ are updated by feature vectors from $\bm{H}_{c}$ through the $\mathtt{distribute}$ operation, while other row vectors remain zero.


Another popular family of pooling methods is node clustering pooling, which constructs an assignment matrix mapping each fine node to its cluster, and then the clusters are treated as new nodes of the coarsened graph. DiffPool (Ying et al. 2018) is among the first attempts to learn a pooling operator end-to-end. In DiffPool, a GNN is trained to compute a soft assignment matrix from the node features and the graph connectivity. Inspired by spectral clustering, MinCut pooling (Bianchi et al. 2020) learns a soft assignment matrix by optimizing a differentiable normalized cut objective. Due to the operations on dense assignment matrices, pooling methods based on learning a clustering operation suffer from high memory cost, making them infeasible for large-scale graphs (Grattarola et al. 2022; Liu et al. 2022).

\subsubsection{AMG pooling}

In this study, we focus on an AMG-inspired pooling method which is sparse and non-trainable for Graph U-Net architectures. Unlike learnable pooling operators, this non-trainable operator does not require additional parameters or optimization objectives, thus avoiding increased complexity in the training process.

Inspired by aggregation-based AMG (Vanek et al. 1996; Muresan and Notay 2008), the main procedure of graph coarsening involves grouping the nodes into $n_c$ nonempty disjoint sets $\left\{ \mathcal{C}_j \right\}_{j=1}^{n_c}$, called aggregates (clusters). Each aggregate corresponds to one node in the coarse graph. The coarsened node feature and adjacency matrices are then represented, respectively, as
\begin{equation}
	\bm{H}_{c} = \bm{P}^T \bm{H}, \quad \bm{A}_{c} = \bm{P}^T \bm{A} \bm{P}.
\end{equation}
where $\bm{P} \in \mathbb{R}^{n \times n_c}$ is a prolongation matrix (cluster assignment matrix), and $\bm{P}^T$ is the restriction matrix in the Galerkin formulation of AMG. The coarsened adjacency matrix $\bm{A}_{c}$ determines the connectivity of aggregates, and its entries can be computed by
\begin{equation}
	(\bm{A}_{c})_{ij} = \sum_{k \in \mathcal{C}_i} \sum_{l \in \mathcal{C}_j} \bm{P}^T_{ik} \bm{A}_{kl} \bm{P}_{lj} = \sum_{k \in \mathcal{C}_i} \sum_{l \in \mathcal{C}_j} \bm{P}_{ki} \bm{A}_{kl} \bm{P}_{lj} 
\end{equation}
We can see that a node pair $(i, j)$ in the coarse graph $\mathcal{G}_c$ are connected if any of the constituent nodes in the aggregates $\mathcal{C}_i$ and $\mathcal{C}_j$ are neighbors in $\mathcal{G}$.

An AMG method requires formation of an appropriate prolongation matrix. Here we rely on a simple prolongator
\begin{equation}
	\bm{P}_{ij} = \begin{cases} 
		1 & i \in \mathcal{C}_j  \\
		0 & \text{otherwise}
	\end{cases}
\end{equation}
which merely depends on the definition of aggregates. Note that $\bm{P}$ is a sparse matrix by formulation and hence the coarsening operations can be implemented efficiently.

\subsubsection{Aggregation methods}
\label{Aggreg_three}

To create the aggregates $\left\{ \mathcal{C}_j \right\}_{j=1}^{n_c}$, we will consider and study three different aggregation methods (note that the aggregation here for graph coarsening should not be confused with the aggregation operation in the message-passing scheme):
\begin{enumerate}[\quad (1)]
	
	\item $\mathtt{VoxelGrid}$ (Simonovsky and Komodakis 2017) is a straightforward clustering strategy, where a regular grid of user-defined size is superimposed on the graph, and all the nodes within the same voxel are grouped into a single cluster.
	
	\item $\mathtt{Graclus}$ (Dhillon et al. 2007) is an efficient graph clustering algorithm that halves the node set, and has seen widespread use in the graph learning domain. It can provide a scalable approximation to spectral clustering, by optimizing the cut-based objectives without eigenvector computation.
	
	\item $\mathtt{Lloyd}$ (Lloyd 1982; Bell 2008; Nytko 2022) is an efficient aggregation method which has $\mathcal{O}(n)$ time complexity. It uses the Bellman-Ford algorithm (Leiserson et al. 1994) to construct aggregates based on an initial seeding. The details regarding the Lloyd method are summarized in Appendix B.
	
\end{enumerate}

Given that our graph learning task targets a PDE system with spatially varying coefficients, it is crucial to integrate heterogeneous information into graph partitioning algorithms. We note that the interface transmissibility $\Upsilon_{ij}$ (cell connectivity), arising from the finite volume discretization, has the physical meaning: larger values of $\Upsilon$ indicate higher fluid flow capability.

Consequently, the weighted adjacency matrix $\bm{A}$—configured with entries $\Upsilon_{ij}$ as edge weights (after proper normalization) that measure the connection strengths between nodes in the mesh graph—can be leveraged to adapt the partitioning algorithms ($\mathtt{Graclus}$ or $\mathtt{Lloyd}$) to the heterogeneous properties of the physical system. By accounting for fine-scale transmissibilities, the graph partitioning procedure is expected to yield better coarse representations within the hierarchy, ultimately leading to higher inference accuracy of Graph U-Net models.

In our implementation, we perform aggregation and construct the information necessary for pooling/unpooling operations at all the coarse levels, through a preprocessing step before model training. The computational procedure for a data sample (trajectory) is given in Algorithm~\ref{alg:Preprocessing_GU}.

\begin{algorithm}
	\caption{Preprocessing for Graph U-Net}
	\label{alg:Preprocessing_GU}
	\KwIn{\\
		$\mathfrak{L}$: \, $\textrm{{\small Number of coarse levels}}$ \\
		$\Upsilon$: \! \! $\textrm{{\small Cell interface transmissibilities }}$ \\ 		
		$\mathcal{E}_0$: \ $\textrm{{\small Edge list of the input mesh graph}}$ ${\mathcal{G}_0}$
		\\
	}
	\KwOut{\\
		$\bm{A}_{\ell}$, $\bm{P}_{\ell}$ \ for $\ell = 1,2,...,\mathfrak{L}$
	}
	\medskip	
	$\bm{A}_0 \leftarrow \mathtt{ConstructA}\left ( \Upsilon, \mathcal{E}_0 \right )$  \hfill $\{\textrm{{\footnotesize construct weighted adjacency matrix}}\}$ \\ 	
	\For{$\ell = 1,2,...,\mathfrak{L}$ \ }{
		$\bm{Agg}_{\ell} \leftarrow \mathtt{Aggregate}\left ( \bm{A}_{\ell-1} \right )$  \hfill $\{\textrm{{\footnotesize perform aggregation}}\}$ \\ 
		$\bm{P}_{\ell} \leftarrow \mathtt{ConstructP}\left ( \bm{Agg}_{\ell} \right )$  \hfill $\{\textrm{{\footnotesize construct prolongation matrix}}\}$ \\ 
		$\bm{A}_{\ell} \leftarrow \bm{P}^T_{\ell} \bm{A}_{\ell-1} \bm{P}_{\ell}$  \hfill $\{\textrm{{\footnotesize coarsened adjacency matrix}}\}$ \\ 
		$\bm{A}_{\ell} \leftarrow \mathtt{Normalize}\left ( \bm{A}_{\ell} \right )$  \hfill $\{\textrm{{\footnotesize normalize entries}}\}$ \\ 		
	}
\end{algorithm}

\subsection{Graph unpooling}

\subsubsection{AMG unpooling}

To upscale a coarsened graph back to its original size, we can simply transpose the AMG pooling operation (Ruge and Stüben 1987; Eliasof and Treister 2020). The node feature matrix at the immediate finer level is obtained using
\begin{equation}
	\label{eq:unpooling}
	\bm{H}_{f} = \bm{P} \bm{H}
\end{equation}
which corresponds to piecewise constant (unsmooth) interpolation.

Several methods were proposed in AMG literature to obtain a better interpolation (Vanek et al. 1996). One popular method is the so-called smoothed aggregation (SA). For the SA algorithm, $\bm{P}$ can be treated as a tentative prolongator, which is then improved by applying a smoothing (damped Jacobi) iteration
\begin{equation}
	\label{eq:smoothed_unpool}
	\widetilde{\bm{P}} = \left( \bm{I} - \omega \bm{D}^{-1} \bm{A} \right) \bm{P}, \quad \omega = \frac{4}{3} \frac{1}{\theta_{max}\left(\bm{D}^{-1} \bm{A}\right)}.
\end{equation}
where $\bm{D} = \mathrm{diag} (\bm{A})$ and $\theta_{max}$ is the maximum eigenvalue. Note that the smoothing operation extends the range of contributions from neighboring nodes. The resultant smoothed unpooling operator $\widetilde{\bm{P}}$ is no longer a binary matrix, meaning that nodes have differing interpolation weights for each aggregate.

\subsubsection{knn unpooling}

Alternatively, we adopt the k-nearest neighbor (knn) interpolation scheme, proposed by Qi et al. (2017), to serve as an unpooling operator. Specifically, for each node located at position $\bm{x}$, we compute its interpolated features $\bm{h}_{f} \in \bm{H}_{f}$ as 
\begin{equation}
	\label{eq:knn_unpooling}
\bm{h}_{f}(\bm{x}) = \frac{\sum_{i=1}^{n_k} w_i(\bm{x}) \, \bm{h}_i}{\sum_{i=1}^{n_k} w_i(\bm{x})}, \quad w_i(\bm{x}) = \frac{1}{d(\bm{x}, \bm{x}_i)^2}.
\end{equation}
where $d(\bm{x}, \bm{x}_i)$ is the spatial distance, $\bm{x}_i$ and $\bm{h}_i$ are the position and feature vector of a node sourced from the current coarse level, and $\left \{ 1,...,n_k \right \}$ denote the $n_k$ nearest nodes to $\bm{x}$. The method allows reconstructing higher-resolution features by weighting the neighbor contributions inversely proportional to the square of their distances.

\subsection{Architecture detail}

The Graph U-Net architecture used in our modeling studies has three levels of coarser graphs (depth = 3). At the zeroth level, the same encoder/decoder as in the single-level architecture (described in Section \ref{onelevel_gat}), are specified for mapping the nodal input/output of mesh graphs. After the encoding step, there are three initial $\mathtt{GATConv}$ layers with ReLU nonlinearities. These layers take both node features and multi-dimensional edge features (transmissibility and relative node positions) to integrate essential local interactions into node representations.

A fixed pooling ratio of 0.2 is specified at every level. After each $\mathtt{Pool}$/$\mathtt{Unpool}$ layer, we apply a Conv block consists of one $\mathtt{GATConv}$ layer, followed by the Batch Normalization $\mathtt{BatchNorm}$ (Ioffe 2015) and ReLU. Here only spatial distance between nodes as edge weight is utilized for $\mathtt{GATConv}$. The sizes of the input, latent and output node features are 8, 32 and 1, respectively.

\section{Model training}

We train the GNN models using the dynamic state pairs $\left ( \bm{U}^{m}; \bm{U}^{m+1} \right )$ from $n_S$ number of simulated rollout trajectories. We supervise a mean squared error (MSE) loss between the predictions $\widehat{\bm{U}}_{\varsigma}^{m+1}$ and their corresponding ground truth labels $\bm{U}_{\varsigma}^{m+1}$ (high-fidelity simulator reference). The loss function is minimized through
\begin{equation}
	\Theta^{*} = \underset{\Theta}{\textrm{argmin}} \, \frac{1}{n_S} \frac{1}{n_t} \sum_{\varsigma=1}^{n_S} \sum_{m=0}^{n_t-1} \left \| \widehat{\bm{U}}_{\varsigma}^{m+1} - \bm{U}_{\varsigma}^{m+1} \right \|_2^2
\end{equation}
where $n_t$ is the number of timesteps (temporal snapshots), and $\bm{U}_{\varsigma}^{m+1}$ denotes either pressure or water saturation of every mesh node, at time $t^{m+1}$, for training sample $\varsigma$. During training, the network weights of GNN are adjusted according to the loss function gradient using back-propagation.

Simulating a complex time-dependent PDE system demands the model to alleviate error accumulation over long rollout trajectories (Sanchez-Gonzalez et al. 2020). Since we only supervise on ground-truth one-step data for training our surrogates, we corrupt the input states $\bm{U}_{\varsigma}^{m}$ with normal noise $N \left ( 0, \sigma \right )$ of zero mean and fixed variance. In such manner, the rollouts of multiple timesteps from a trained model become robust to their own noisy, previous predictions as input. The scales of added noises are $(\sigma_p = 0.01, \sigma_s = 0.02)$ and $(\sigma_p = 0.001, \sigma_s = 0.02)$ for the pressure and saturation model, respectively.

Our implementation relies on PyAMG (Bell et al. 2022) for AMG routines, along with PyTorch (Paszke et al. 2019) and the PyTorch Geometric (PyG) library (Fey and Lenssen 2019) for developing and training GNNs.

\section{Surrogate model evaluations}

We evaluate prediction performance of the surrogate models and their generalization capabilities on unseen configurations.

\begin{figure}[!htb]
	\centering
	\subfloat[Permeability (log)]{
		\includegraphics[scale=0.35]{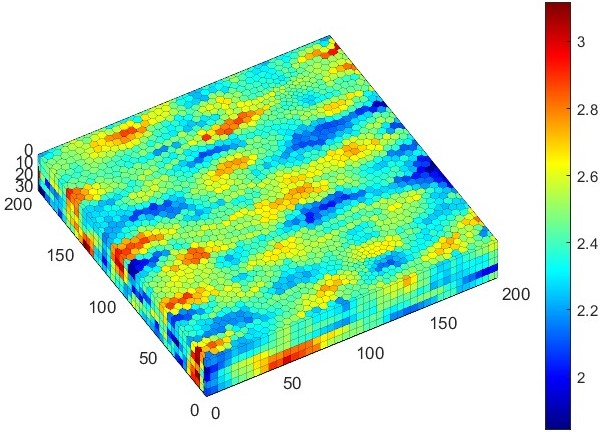}}
	\quad
	\subfloat[Mesh with the wells]{
		\includegraphics[scale=0.35]{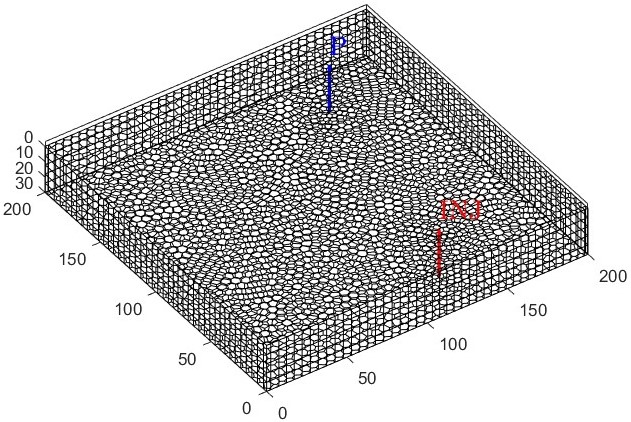}}
	\caption{Mesh and permeability (md) field of an example model case.}
	\label{fig:mesh_rock_3D}
\end{figure}

We consider three-dimensional layered reservoir models containing two wells (one injector and one producer) controlled by constant bottom-hole pressure (BHP). Voronoi diagrams (Yan et al. 2013) are used to create the so-called perpendicular bisector (PEBI) meshes (unstructured polyhedral). The mesh and the heterogeneous rock field of an example model case (with $n = 9690$ mesh cells) is plotted in \textbf{Fig.~\ref{fig:mesh_rock_3D}}. The injector and producer are treated as vertical line source and sink, respectively. No-flow condition $\bm{v}_l \cdot \bm{n} = 0$ is specified at the external boundaries $\partial \Omega$.

The model setup is summarized in Table~\ref{tab:specification_m}. Total simulation time is 190 days, with a fixed 10 days timestep size (19 timesteps). The migration of multi-phase fluid is governed by the complex interplay of viscous, capillary, and gravity forces. Quadratic relative permeabilities are used. The capillary pressure is computed through a simple linear function as
\begin{equation}
	p_{ca}(s^{\ast}) = p_e \, s^{\ast},
\end{equation}
where $p_e = 7.25 \, \textrm{psi}$ is the capillary entry pressure, and $s^{\ast}$ is the normalized (effective) water saturation
\begin{equation}
	s^{\ast} = \frac{s_w - s_{wr}}{1 - s_{wr}}.
\end{equation}

\begin{table}[!htb]
	\centering
	\caption{Setup of the base model}
	\label{tab:specification_m}
	\begin{tabular}{|c|c|c|}
		\hline
		Parameter                  &  Value          & Unit   \\ \hline
		Model sizes (x, y, z)      &  200, 200, 30    & m      \\ \hline
		Initial pressure           &  2000           & psi    \\ \hline
		Initial water saturation   &  0.01           & (-)       \\ \hline
		Water density              &  1000            & $\textrm{kg}/\textrm{m}^3$  \\ \hline
		Non-wetting phase reference density  &  800             & $\textrm{kg}/\textrm{m}^3$  \\ \hline
		Water viscosity            &  1.0            & cP     \\ \hline
		Non-wetting phase viscosity              &  2.0            & cP     \\ \hline
		Rock porosity              &  0.2           & (-)  \\ \hline
		Rock compressibility       &  1e-8           & 1/bar  \\ \hline
		Non-wetting phase compressibility        &  1e-4           & 1/bar  \\ \hline
		Production BHP             &  1800           & psi  \\ \hline
		Injection BHP              &  2200           & psi  \\ \hline
		Total simulation time      &  190            & day     \\ \hline
		Timestep size              &  10             & day     \\ \hline
	\end{tabular}
\end{table}

\begin{figure}[!htb]
	\centering
	\includegraphics[scale=0.44]{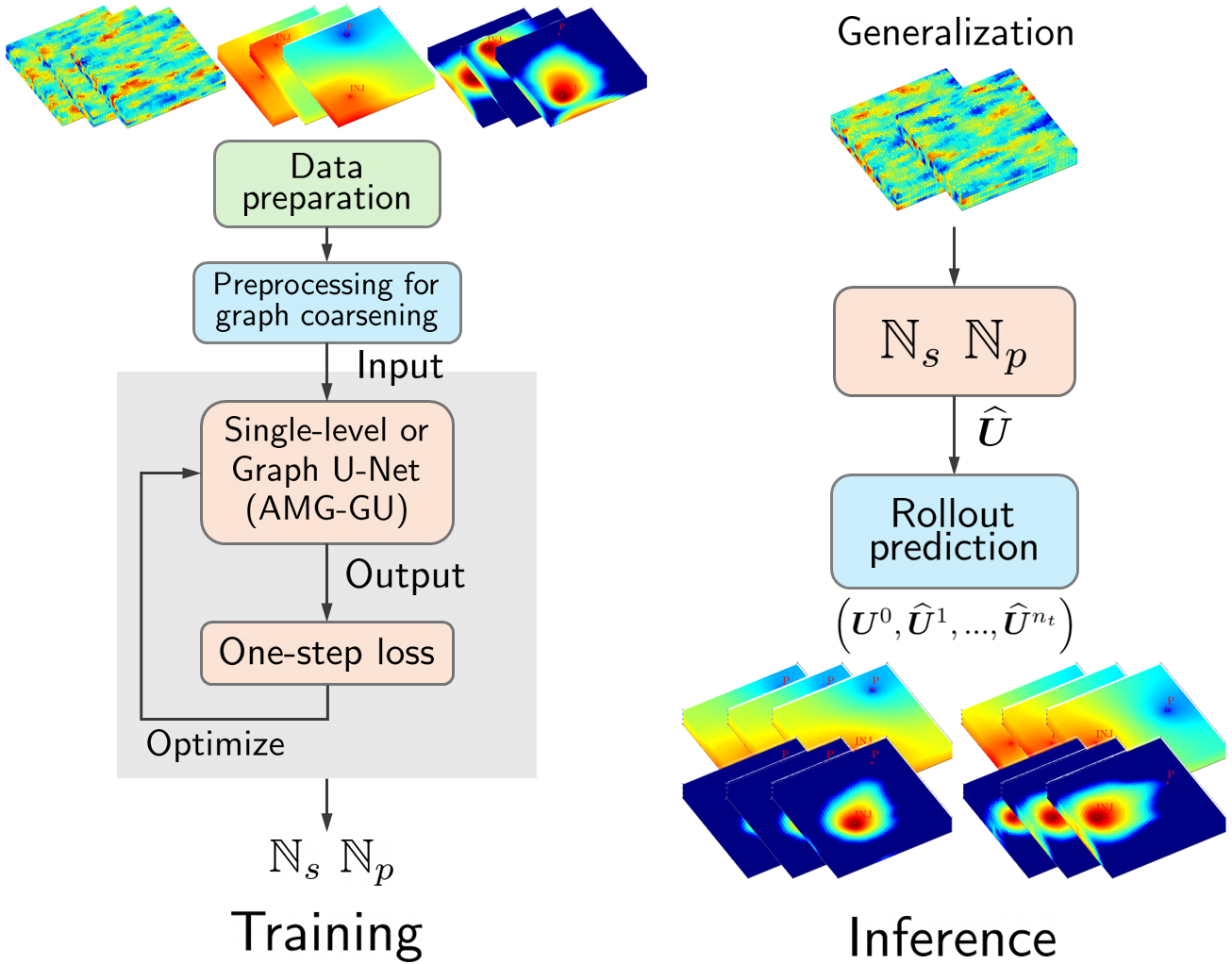}
	\caption{Schematic of the training and inference stages for surrogate modelling.}
	\label{fig:trainInference}
\end{figure}

The overall surrogate modelling procedure is shown in \textbf{Fig.~\ref{fig:trainInference}}. The open-source MATLAB Reservoir Simulation Toolbox (MRST) is employed to generate the datasets (Lie et al. 2012; Lie and Møyner 2021). There are a total of 200 high-fidelity simulation runs ($\textrm{200 trajectories} \, \times \textrm{19 timesteps} = \textrm{3800 snapshots}$) as training data with random well locations and rock properties. The realizations of heterogeneous permeability fields are generated using a Gaussian distribution. Near-well mesh refinement is applied to achieve higher resolution for large pressure gradients and flow rates around wells. This will also introduce perturbation to the realizations, so that each sample (trajectory) will have a different PEBI mesh.

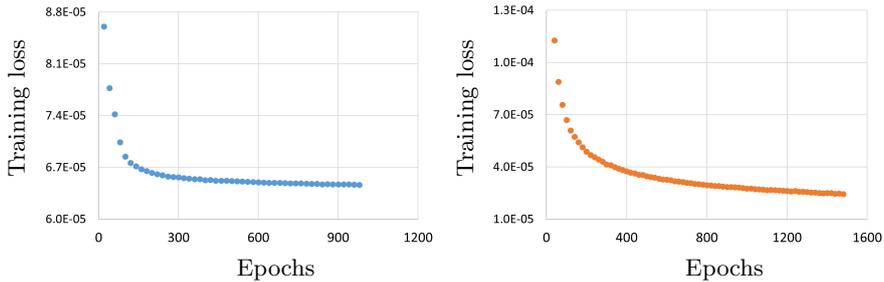
\begin{figure}[!htb]
	\centering
	\begin{minipage}{0.42\textwidth}
		\centering
		\begin{tikzpicture}
			\node (img)  {\includegraphics[scale=0.45]{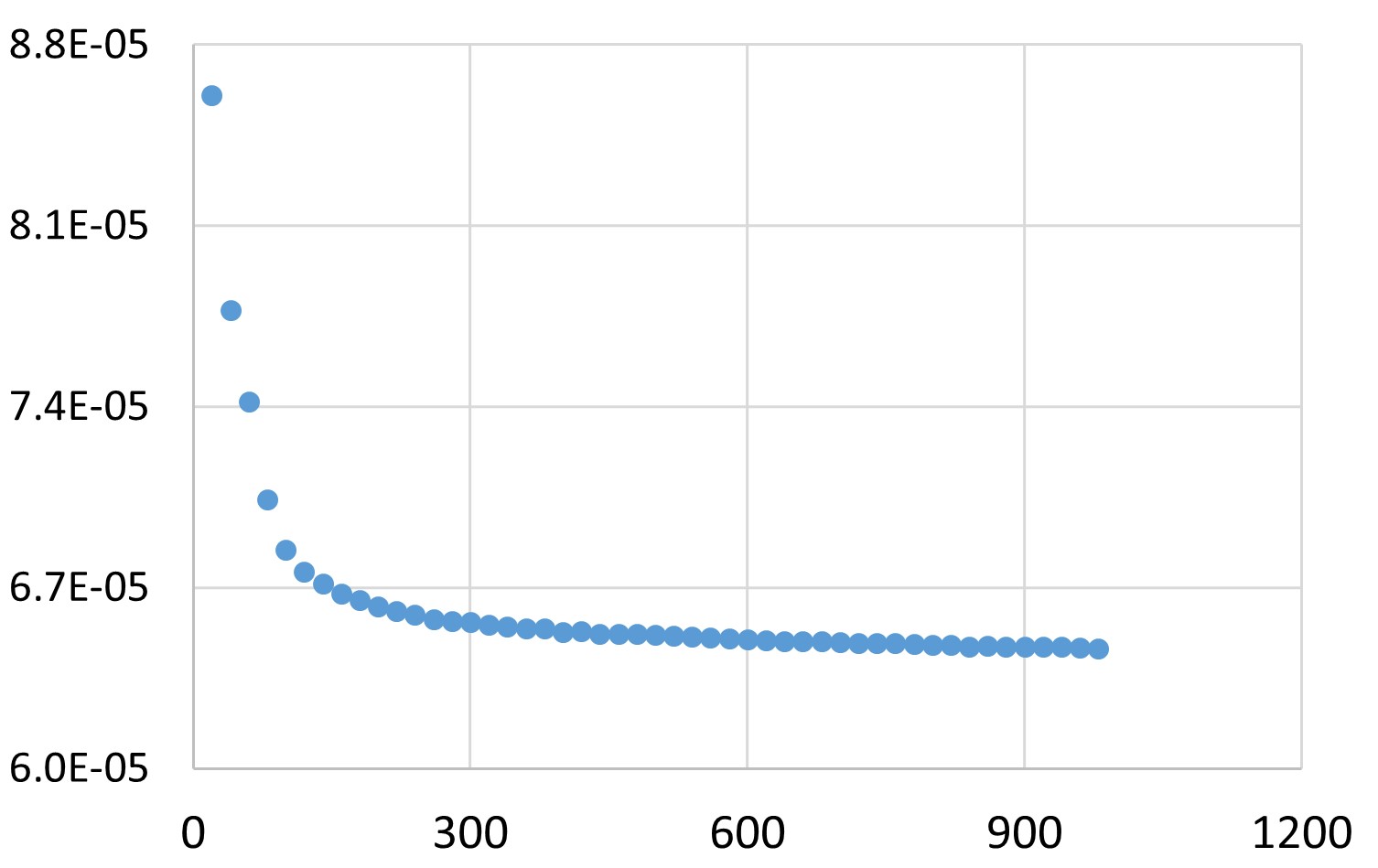}};
			\node[below=of img, node distance=0cm, xshift=0.45cm, yshift=1.1cm, font=\small] {Epochs};
			\node[left=of img, node distance=0cm, rotate=90, anchor=center, xshift=0.2cm, yshift=-0.8cm, font=\small] {Training loss};
		\end{tikzpicture}
	\end{minipage}
	\
	\begin{minipage}{0.42\textwidth}
		\centering
		\begin{tikzpicture}
			\node (img)  {\includegraphics[scale=0.45]{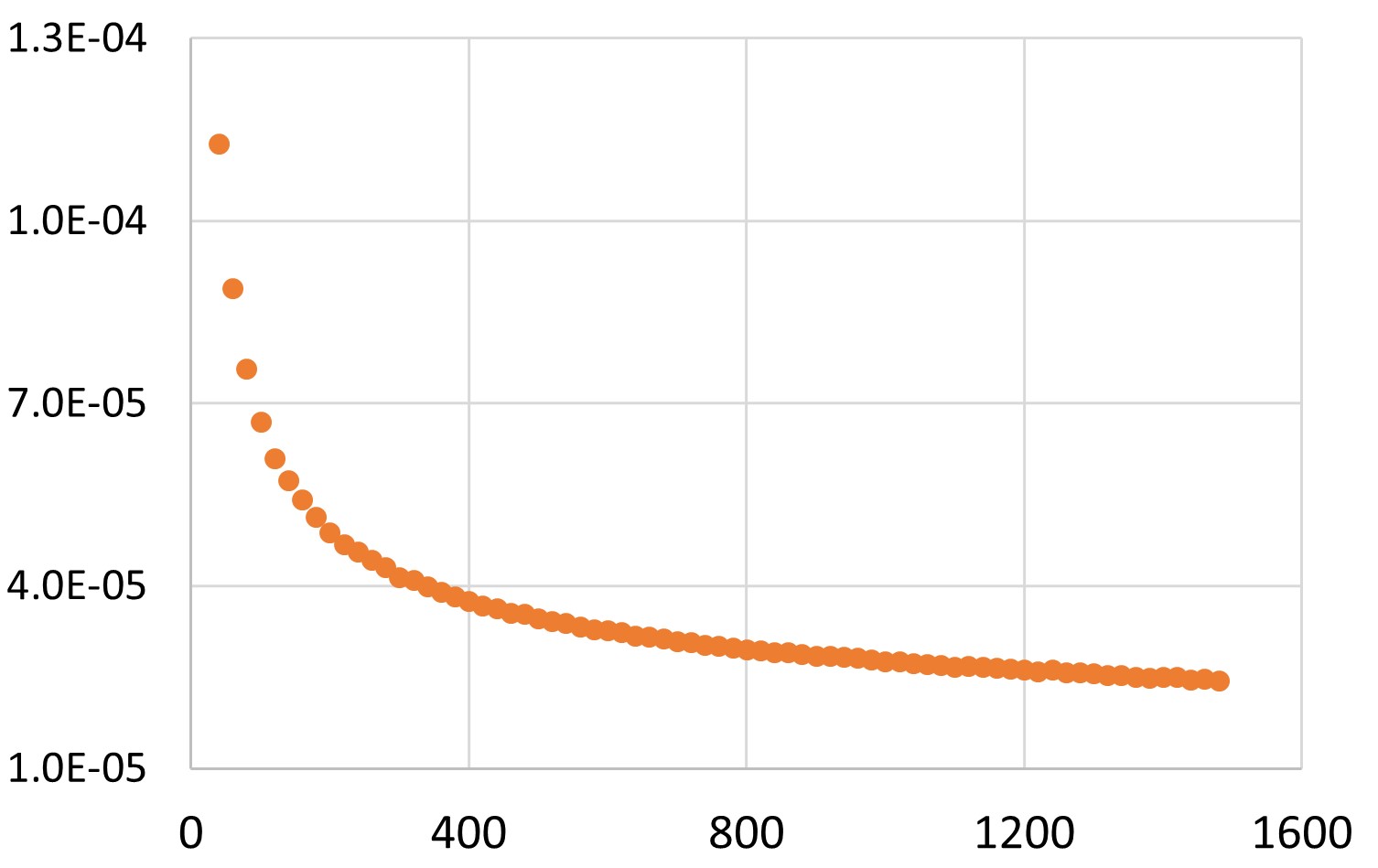}};
			\node[below=of img, node distance=0cm, xshift=0.45cm, yshift=1.1cm, font=\small] {Epochs};
			\node[left=of img, node distance=0cm, rotate=90, anchor=center, xshift=0.2cm, yshift=-0.8cm, font=\small] {Training loss};
		\end{tikzpicture}
	\end{minipage}
	
	\caption{Training loss (MSE) curves of the pressure (left) and saturation (right) models.}
	\label{fig:training_loss}
\end{figure}

The GNN models have been trained on a NVIDIA GeForce RTX 4090 GPU using the Adam optimizer (Kingma and Ba 2014) with learning rate 1e-4. The training loss (MSE) curves are plotted in \textbf{Fig.~\ref{fig:training_loss}}. It takes approximately 4 and 8 hours to train the pressure and saturation models, respectively. The trained surrogates can predict a trajectory (with $\sim 10000$ mesh nodes) in 0.1 seconds, achieving a substantial reduction of computational time compared to the reference simulator. A single high-fidelity simulation run requires about 16 seconds on an Intel Core i7-12800HX CPU.

We generate a total of 20 testing samples, and will provide qualitative as well as quantitative comparisons of the solutions (pressure and water saturation) between the surrogate (prediction) and high-fidelity (ground truth) simulators.

\subsection{Aggregation methods}

\begin{figure}[!htb]
	\centering
	\subfloat[$\mathtt{VoxelGrid}$ ($n_c = 288$)]{
		\includegraphics[scale=0.3]{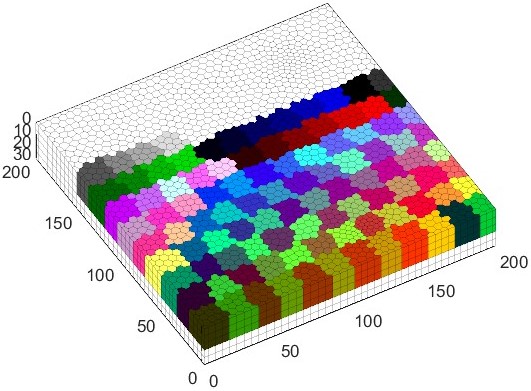}}
	\,
	\subfloat[$\mathtt{Graclus}$ ($n_c = 2725$)]{
		\includegraphics[scale=0.3]{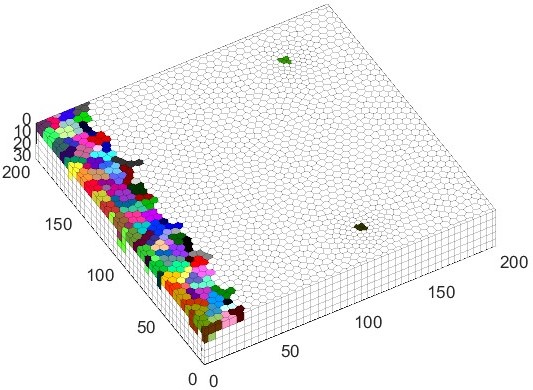}}
	\,
    \subfloat[$\mathtt{Lloyd}$ ($n_c = 290$)]{
	\includegraphics[scale=0.3]{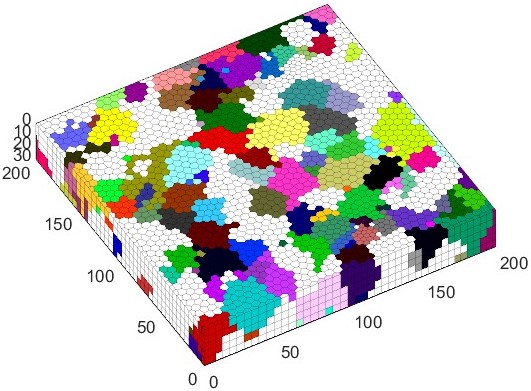}}
	\caption{Aggregations from the three methods. $n_c$ is the number of aggregates (coarse nodes). Portions of the aggregates are ignored for better visualization.}
	\label{fig:three_clusters}
\end{figure}

We demonstrate an application of the three aggregation methods mentioned in Section~\ref{Aggreg_three} by coarsening the example mesh (\textbf{Fig.~\ref{fig:mesh_rock_3D}}). A relative aggressive pooling ratio (0.03 for $\mathtt{VoxelGrid}$ and $\mathtt{Lloyd}$) is specified, and the resulting aggregations are plotted in \textbf{Fig.~\ref{fig:three_clusters}}. Note that $\mathtt{Graclus}$ lacks precise control of the number of aggregates, and thus recursive executions are necessary to reach the desired coarsened graph.

We can see that $\mathtt{VoxelGrid}$ produces a relatively regular and uniform aggregation, while completely ignoring the heterogeneous property of the underlying physical system. By comparison, the aggregates from $\mathtt{Graclus}$ and $\mathtt{Lloyd}$ are more irregular and have larger size variations. As previously discussed, these two methods can adapt to fine-scale heterogeneities through a weighted adjacency matrix measuring strengths of the cell connections.

\subsection{Comparisons to baseline models}

We evaluate our Graph U-Net architecture by considering two baselines: (1) \textbf{Single-level GAT}: the single-level GAT architecture described in Section \ref{onelevel_gat}. (2) \textbf{TopK-GU}: Graph U-Net with the TopK pooling from Gao and Ji (2021). Regarding the AMG-inspired Graph U-Net model (referred to as \textbf{AMG-GU}), here we choose $\mathtt{Lloyd}$ for aggregation and the unsmooth (piecewise constant) prolongator (Eq.~\ref{eq:unpooling}) for unpooling.

\begin{figure}[!htb]
	\centering
	\subfloat[$\mathsf{Case\ 1}$]{
		\includegraphics[scale=0.26]{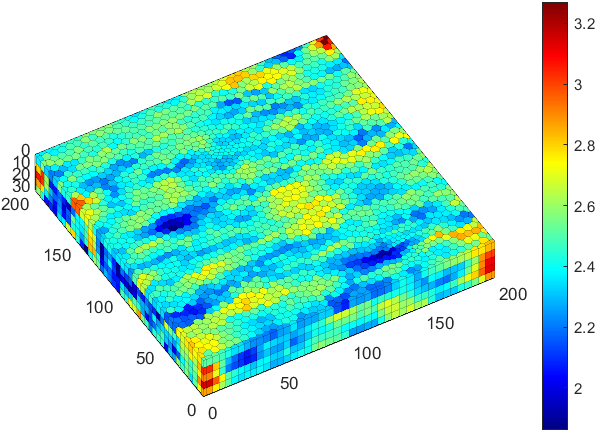}}
	\ \
	\subfloat[$\mathsf{Case\ 2}$]{
		\includegraphics[scale=0.26]{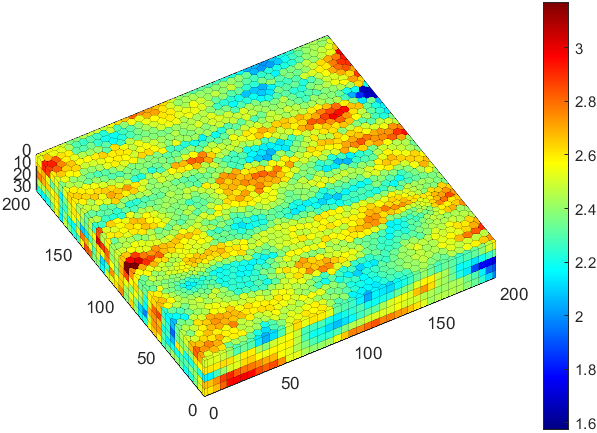}}
	\ \
	\subfloat[$\mathsf{Case\ 3}$]{
		\includegraphics[scale=0.26]{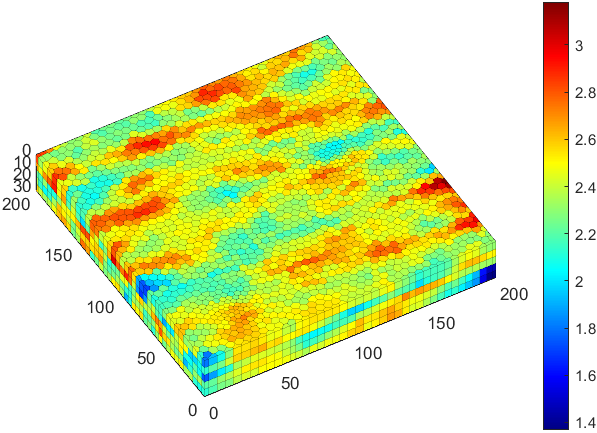}}
	\caption{Permeability (log) fields of the three testing cases.}
	\label{fig:sec1_3cases_rock}
\end{figure}
\begin{figure}[!htb]
	\centering
	\subfloat[Pressure (psi)]{
		\includegraphics[scale=0.35]{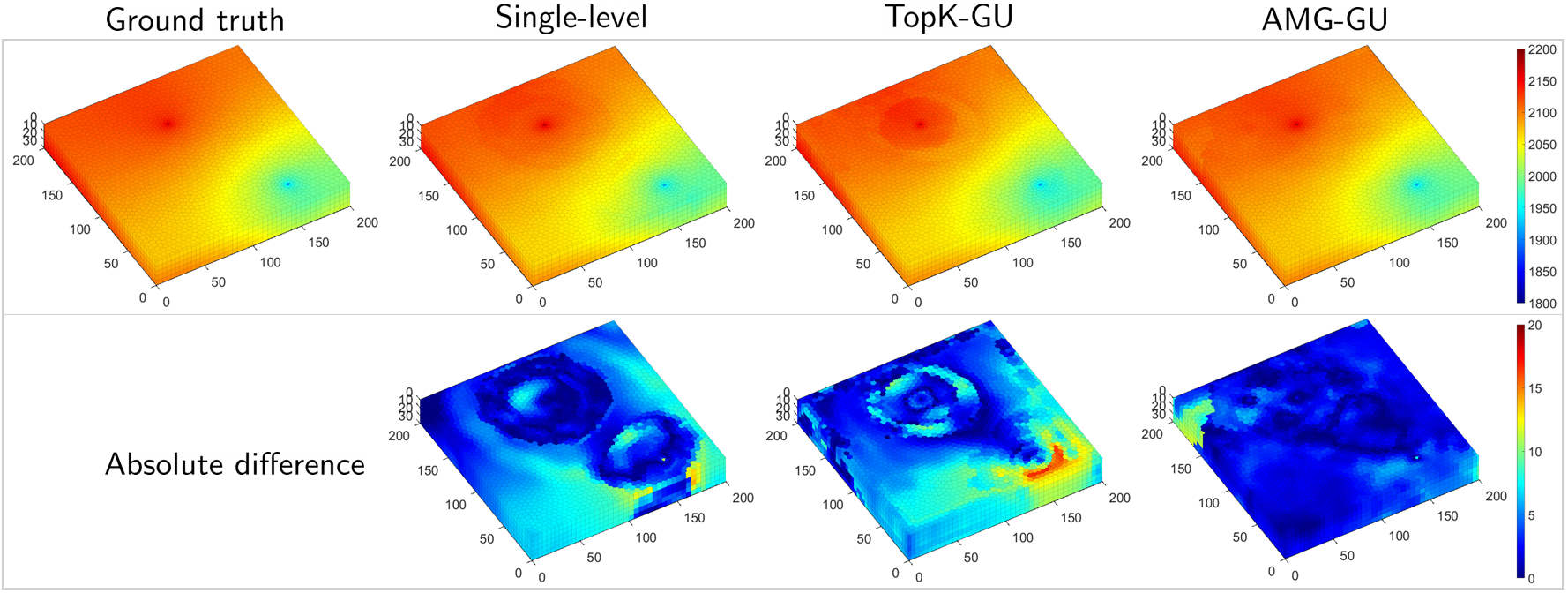}}
	\
	\subfloat[Saturation]{
		\includegraphics[scale=0.35]{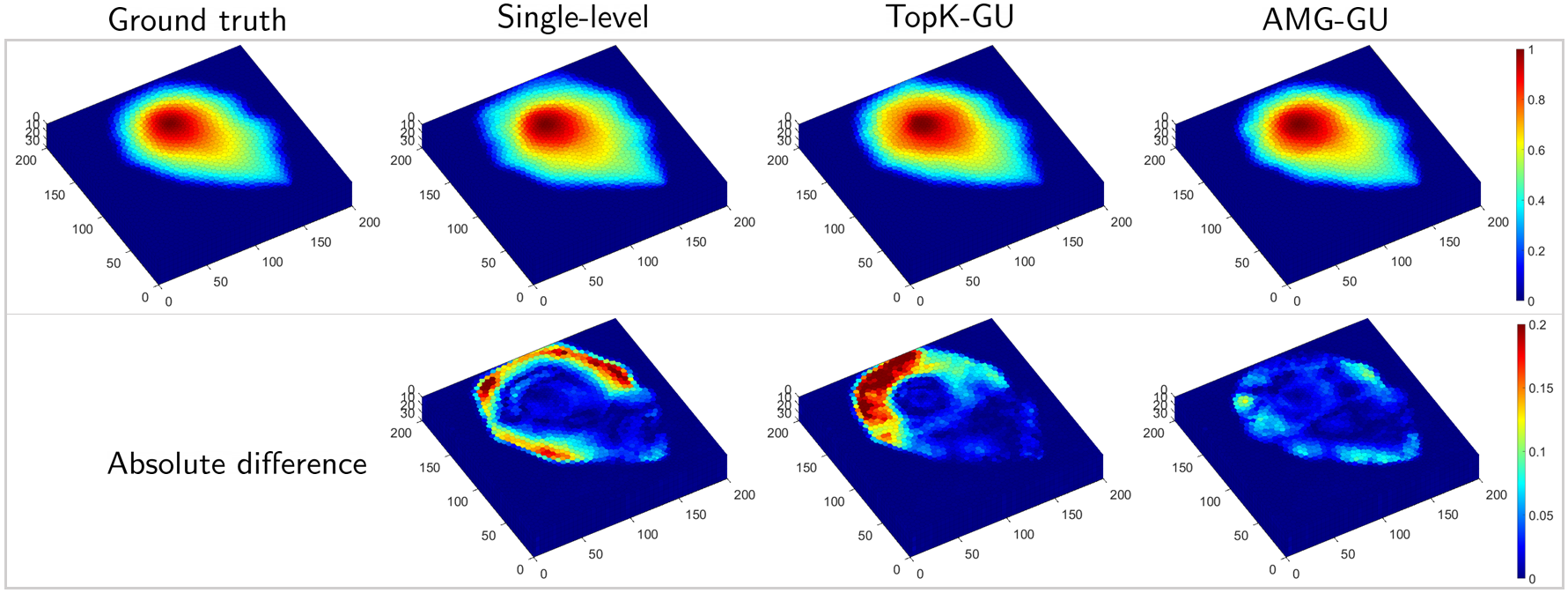}}
	\caption{Solution profiles of $\mathsf{Case\ 1}$.}
	\label{fig:sec1_1_ps}
\end{figure}
\begin{figure}[!htb]
	\centering
	\subfloat[Pressure (psi)]{
		\includegraphics[scale=0.35]{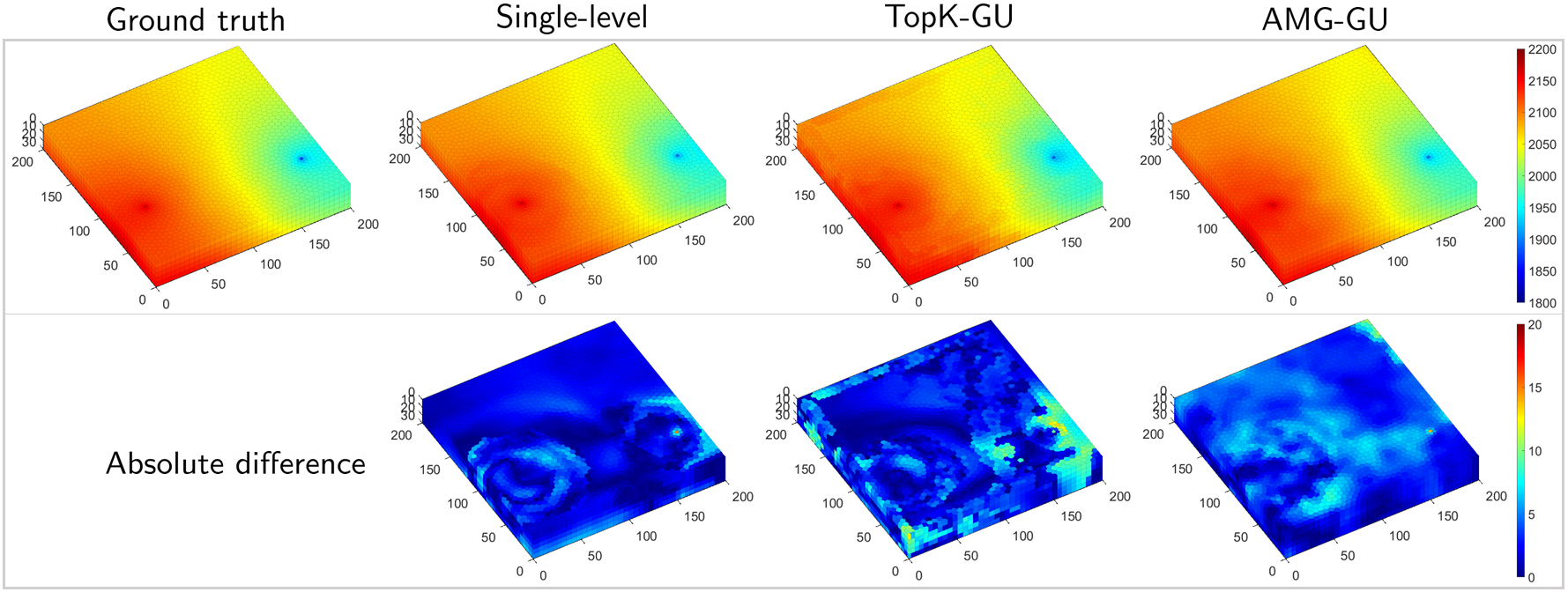}}
	\
	\subfloat[Saturation]{
		\includegraphics[scale=0.35]{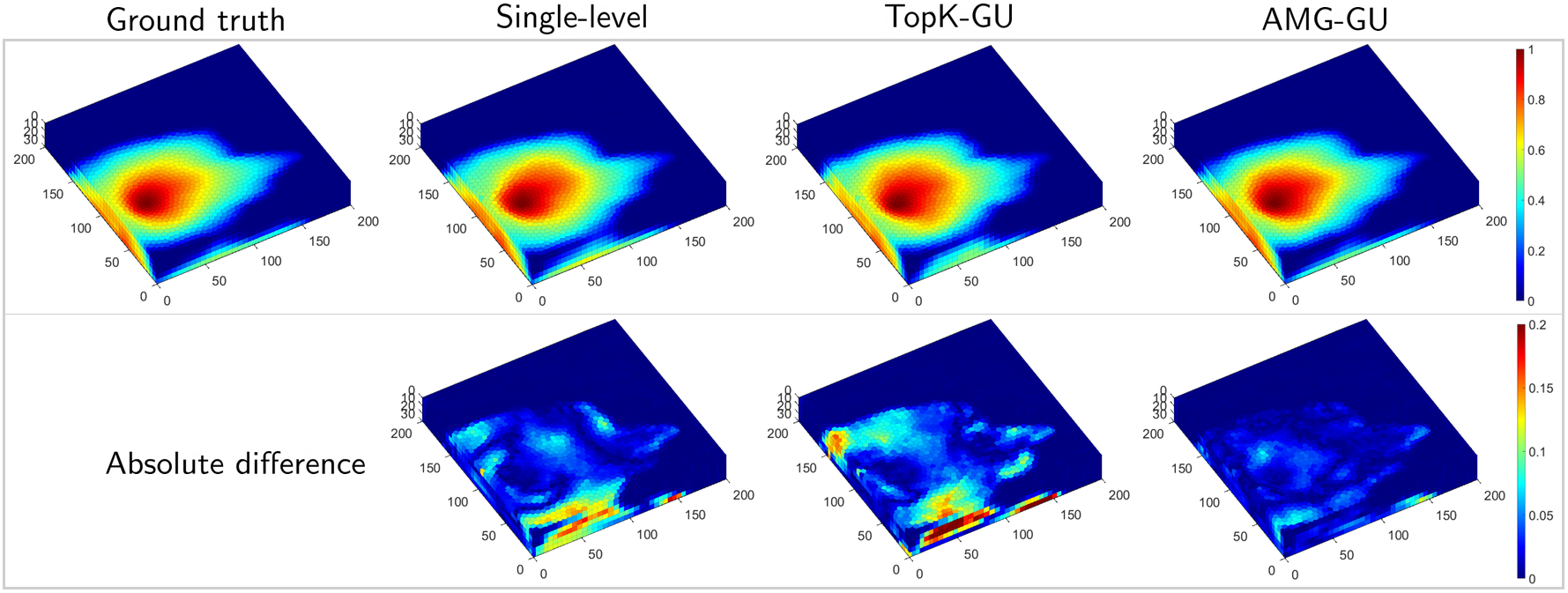}}
	\caption{Solution profiles of $\mathsf{Case\ 2}$.}
	\label{fig:sec1_2_ps}
\end{figure}
\begin{figure}[!htb]
	\centering
	\subfloat[Pressure (psi)]{
		\includegraphics[scale=0.35]{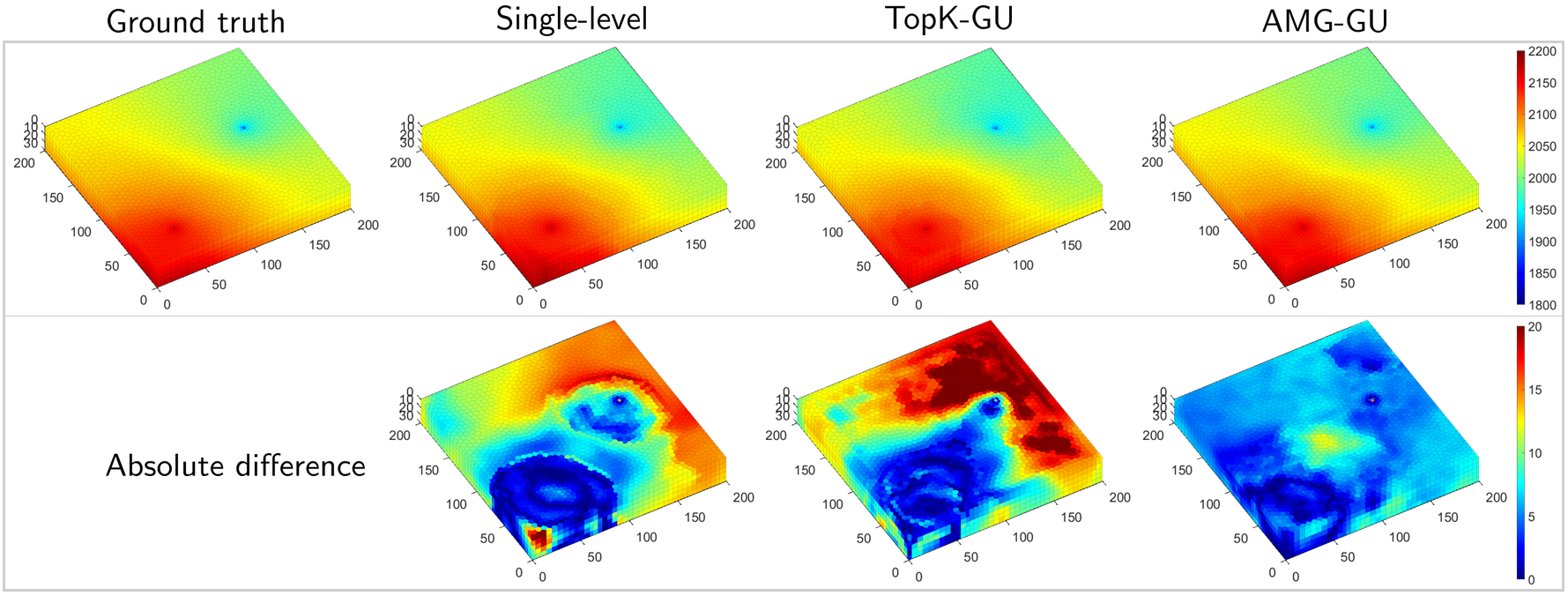}}
	\
	\subfloat[Saturation]{
		\includegraphics[scale=0.35]{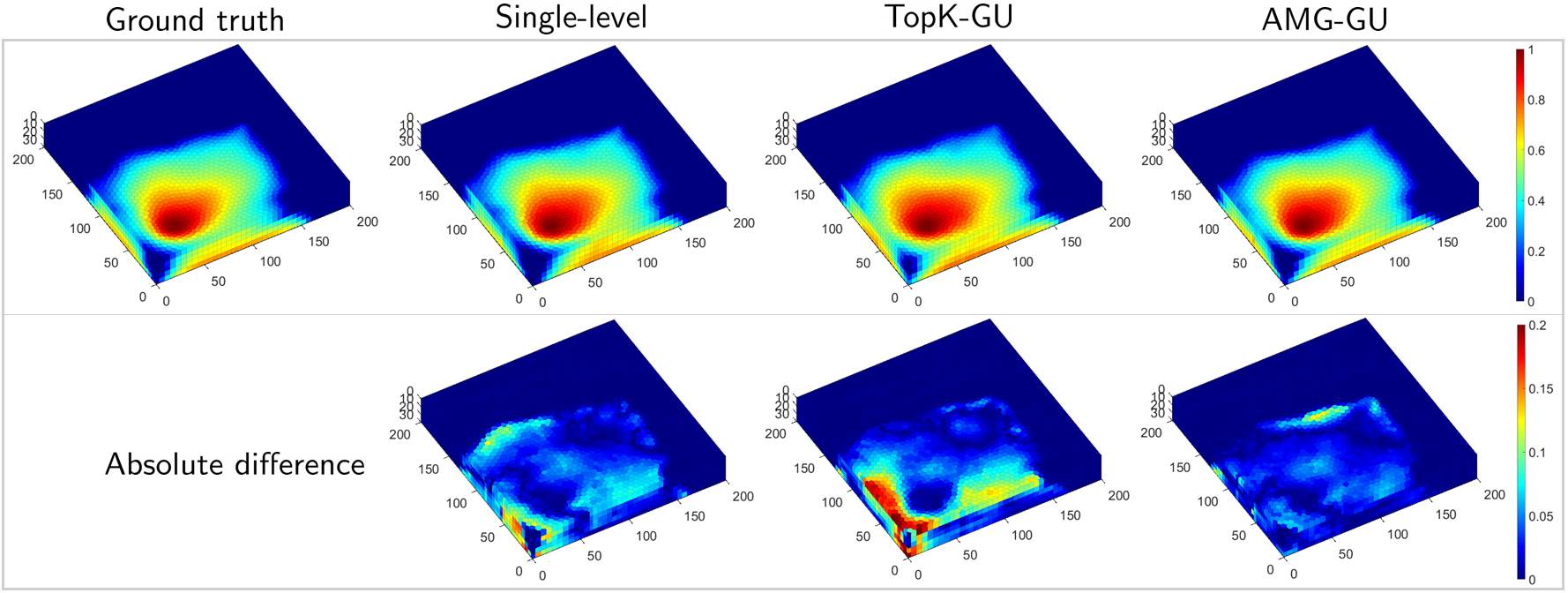}}
	\caption{Solution profiles of $\mathsf{Case\ 3}$.}
	\label{fig:sec1_3_ps}
\end{figure}

We first present the predictions of three representative cases from the testing set. The permeability fields of the three cases are shown in \textbf{Fig.~\ref{fig:sec1_3cases_rock}}. The pressure and water saturation solution profiles are shown in \textbf{Fig.~\ref{fig:sec1_1_ps}}, \textbf{Fig.~\ref{fig:sec1_2_ps}} and \textbf{Fig.~\ref{fig:sec1_3_ps}}, respectively. Only qualitative and quantitative comparisons of the solutions at the end timestep between the surrogate and high-fidelity simulators are analyzed, since the final snapshots of a trajectory should exhibit the biggest accumulated errors.


From the results we can see that the well configurations have a dominant influence on the pressure profiles. The prediction errors from the three surrogate models are distributed across the entire domain because of the parabolic nature of the pressure component. While the two baseline models capture overall shape and structure of the flow patterns, certain unphysical values (non-monotone and non-smooth) that violate the underlying parabolic PDE can be clearly observed. In contrast, AMG-GU gives monotone and smooth pressure solutions for all the three test cases, except at some small regions near the domain boundaries.


The saturation distributions are strongly impacted by the well locations and heterogeneous properties. We can see that the predictions based on the two baselines exhibit large local errors that are evident near the domain boundaries and fluid fronts. In comparison, AMG-GU leads to much smaller discrepancies, reproducing both the shapes and heterogeneous details of sharp fronts with reasonable accuracy. 

This is because of the more accurate pressure approximations obtained by the AMG-inspired Graph U-Net. Note that the solution quality of the parabolic component can have a vital influence on the saturation prediction during the evolution of the coupled multi-phase system (Eq.~\ref{eq:cm_PS}). Accurate pressure solutions will result in a flux field of good quality, which in turn will lead to better saturation solutions (see Eq.~\ref{eq:fff_vw}), even though flux field does not appear explicitly as input for the surrogate $\mathbb{N}_s$. In summary, our AMG-GU model is capable of making qualitative predictions of fluid distribution with sufficient accuracy under the three test cases.

\begin{figure}[!htb]
	\centering
	\subfloat[Pressure (psi)]{
		\includegraphics[scale=0.41]{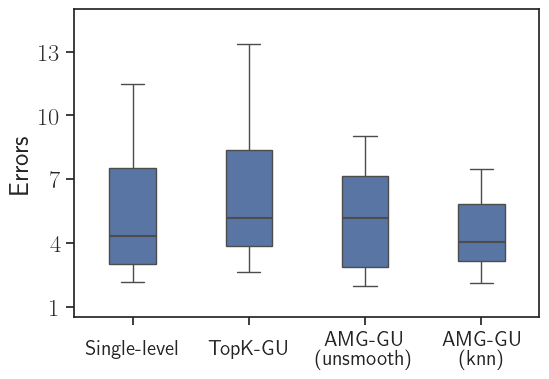}}
	\ \ 
	\subfloat[Saturation]{
		\includegraphics[scale=0.41]{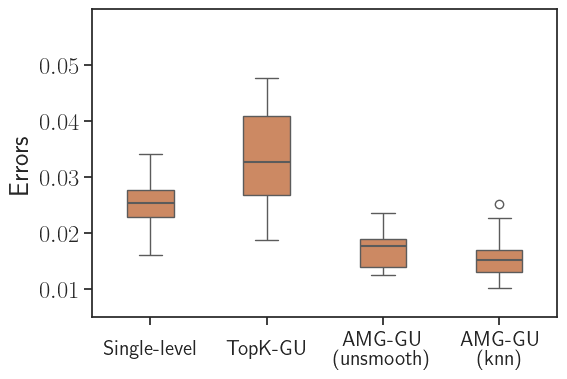}}
	\caption{Boxplots of the mean absolute pressure and saturation errors for each of the testing cases (comparisons to the baselines).}
	\label{fig:sec1_box_ps}
\end{figure}

To quantitatively assess the predicted pressure and saturation snapshots, we use a metric for individual testing samples. The mean absolute error (MAE) is given as
\begin{equation}
	\delta^{A} = \frac{1}{n} \sum_{i=1}^{n} \left \| \widehat{u}_i - u_i \right \|
\end{equation}
where $n$ is the number of mesh cells in a snapshot. The boxplots of the error metrics $\delta_p$ and $\delta_s$ for each of the 20 testing samples are shown in \textbf{Fig.~\ref{fig:sec1_box_ps}}.


Here we additionally add the results based on the AMG-GU model using the knn unpooling operator (Eq.~\ref{eq:knn_unpooling}) for comparison. The mean absolute pressure and saturation errors from the 4 surrogates over all the testing samples are \{5.4, 6.3, 5.1, 4.4\} psi and \{0.025, 0.033, 0.017, 0.015\}, respectively. Note that the maximum value differences of pressure and saturation are respectively 400 psi and 1.0 in the simulation data. We can clearly see that the two variants of AMG-GU not only achieve lower mean errors than the Single-level and TopK-GU baselines, but also exhibit a narrower spread of errors (especially saturation). AMG-GU(knn) presents the best prediction performance, showing the benefit of using the smoother interpolation for unpooling. Overall, the low mean errors over the testing samples demonstrate that the AMG-GU models can accurately approximate the dynamical states. This inference capability is highly beneficial for a surrogate in the context of uncertainty quantification.


It is worth noting that TopK-GU causes significantly bigger discrepancies than the other models (even Single-level), indicating the detrimental impact of its pooling operation on model performance. This test result is consistent with some existing findings from the literature regarding TopK performance under various node and graph classification tasks. One probable explanation is that the node-drop mechanism leads to loss of node and connectivity information in coarsened graphs.

\subsection{Effects of aggregation and unpooling}

\begin{figure}[!htb]
	\centering
	\subfloat[Pressure (psi)]{
		\includegraphics[scale=0.44]{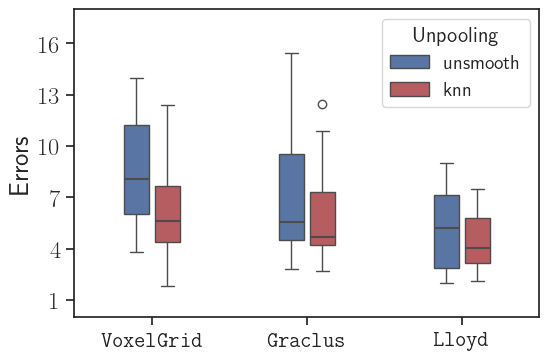}}
	\ \ 
	\subfloat[Saturation]{
		\includegraphics[scale=0.44]{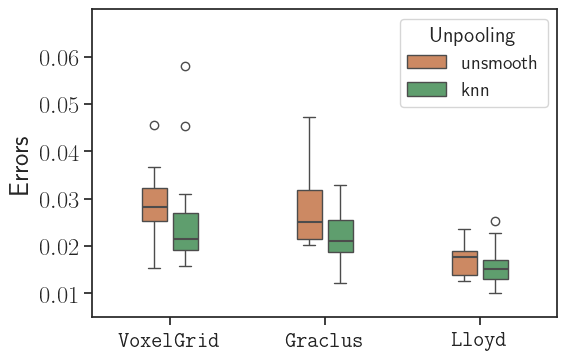}}
	\caption{Boxplots of the mean absolute pressure and saturation errors for each of the testing cases (with different aggregation and unpooling methods in AMG-GU).}
	\label{fig:sec2_box_ps}
\end{figure}

We conduct sensitivity studies on different aggregation and unpooling methods for the AMG-inspired Graph U-Net architecture (AMG-GU). The boxplots of the error metrics $\delta_p$ and $\delta_s$ for each of the 20 testing samples are shown in \textbf{Fig.~\ref{fig:sec2_box_ps}}.

As can be seen, the variant with $\mathtt{Lloyd}$ significantly outperforms $\mathtt{VoxelGrid}$ and $\mathtt{Graclus}$, highlighting importance of the aggregation choice for Graph U-Net. In certain test cases, the pressure and saturation approximations based on $\mathtt{VoxelGrid}$ and $\mathtt{Graclus}$ exhibit quite large deviations, which suggest their lack of good generalization ability to unseen model configurations. As discussed in Section~\ref{Aggreg_three}, the $\mathtt{Lloyd}$ algorithm adapting to fine-scale heterogeneities can form high-quality aggregation that leads to desirable generalizability of AMG-GU surrogates for learning complex fluid dynamics on unstructured meshes. We also observe that the choice of the knn scheme yields consistently lower errors for all the aggregations, confirming that a smoother interpolation is beneficial for unpooling.

\begin{figure}[!htb]
	\centering
	\subfloat[Pressure (psi)]{
		\includegraphics[scale=0.35]{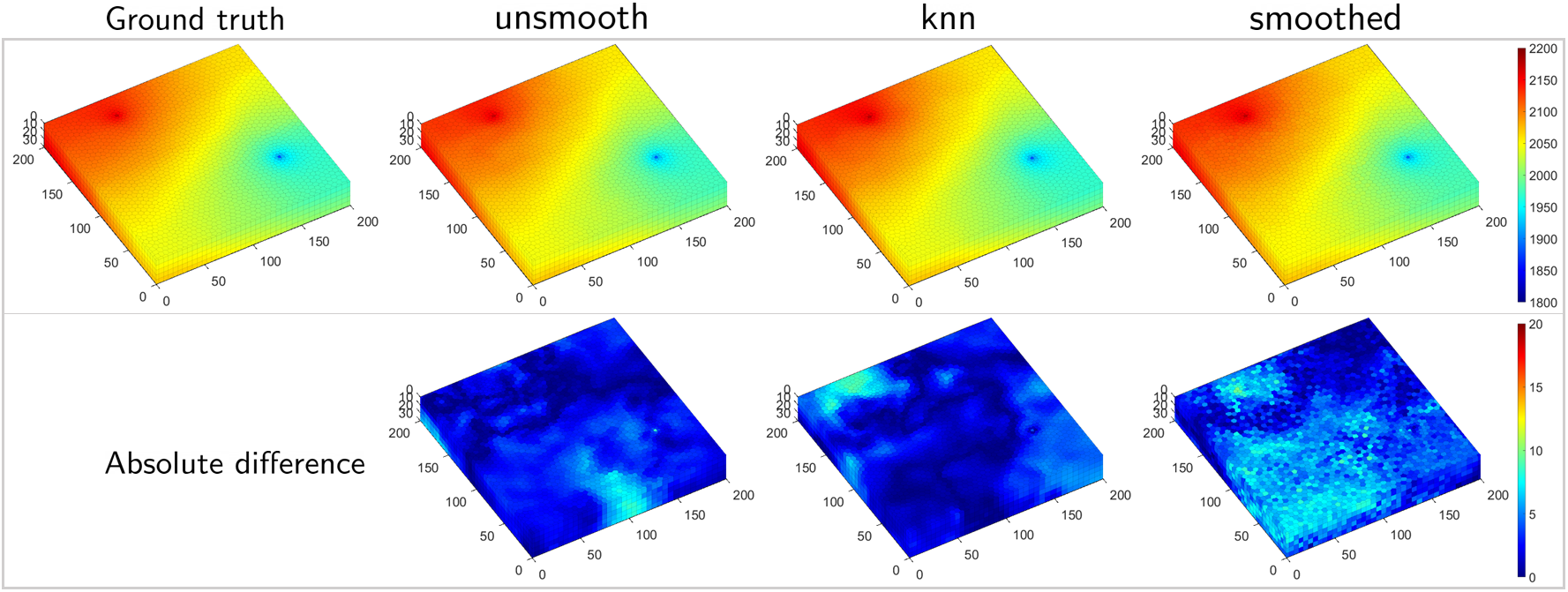}}
	\
	\subfloat[Saturation]{
		\includegraphics[scale=0.35]{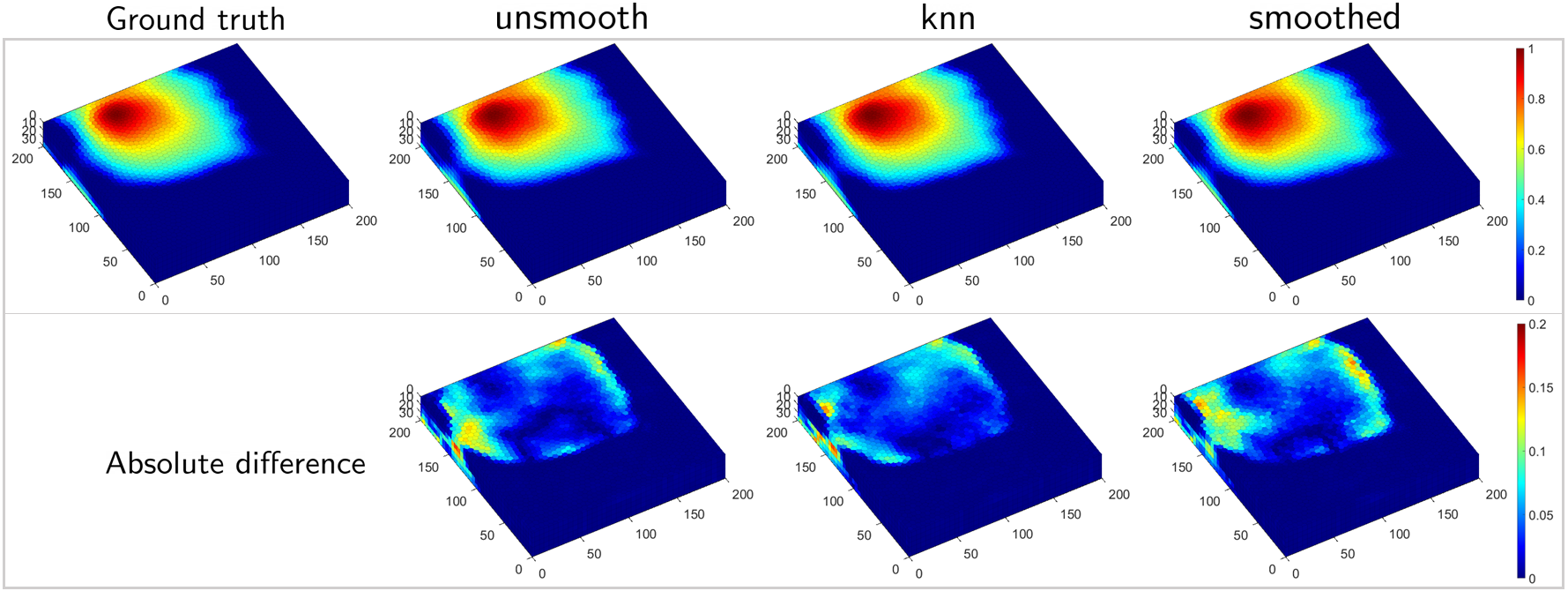}}
	\caption{Solution profiles of $\mathsf{Case\ 4}$ based on the different choices of unpooling operators for AMG-GU with the $\mathtt{Lloyd}$ aggregation.}
	\label{fig:sec2_4_ps}
\end{figure}

Additionally we examine the smoothed prolongator (Eq.~\ref{eq:smoothed_unpool}) in terms of the qualitative results of a representative test case. The comparisons of the pressure and saturation profiles based on the different choices of unpooling operators (using the $\mathtt{Lloyd}$ aggregation) are shown in \textbf{Fig.~\ref{fig:sec2_4_ps}}. Again we can see that the two AMG-GU variants (unsmooth and knn) predict the dynamical states with high accuracy. Note that some smearing of saturation are visible near the boundaries and fluid fronts.

Compared to the reference pressure solution, the profile given by the model with the smoothed unpooling suffers from non-monotonicities and oscillations. This undesirable behavior indicates that the smoothed prolongator does not achieve stable interpolation within the AMG-GU framework. The issue of unphysical solutions needs to be investigated deeper, in order to develop better interpolation schemes for unpooling.

\subsection{Effects of training dataset}

\begin{figure}[!htb]
	\centering
	\subfloat[Permeability and ground truth (pressure and saturation) fields]{
	\includegraphics[scale=0.33]{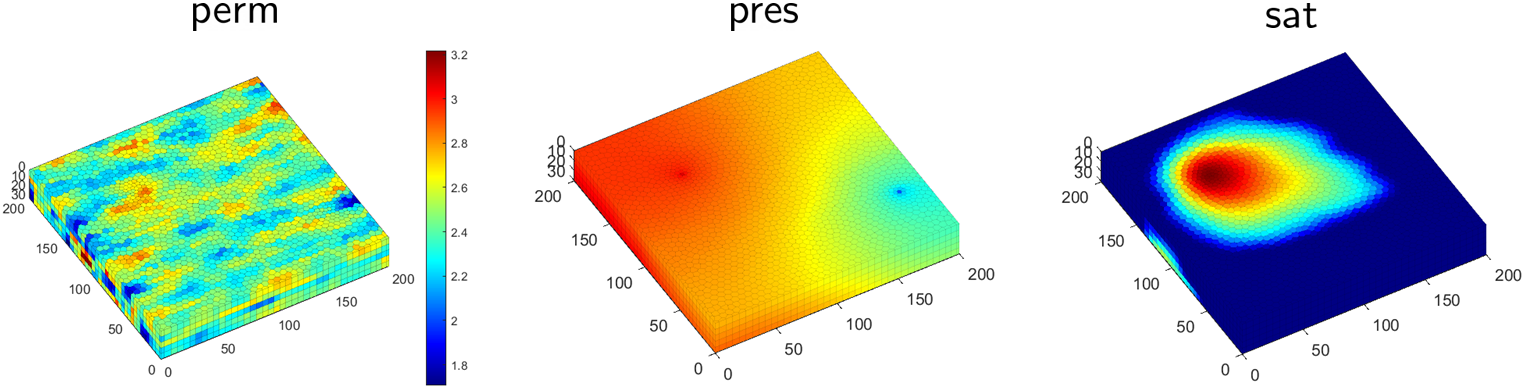}}
	\
	\subfloat[Pressure (psi)]{
		\includegraphics[scale=0.35]{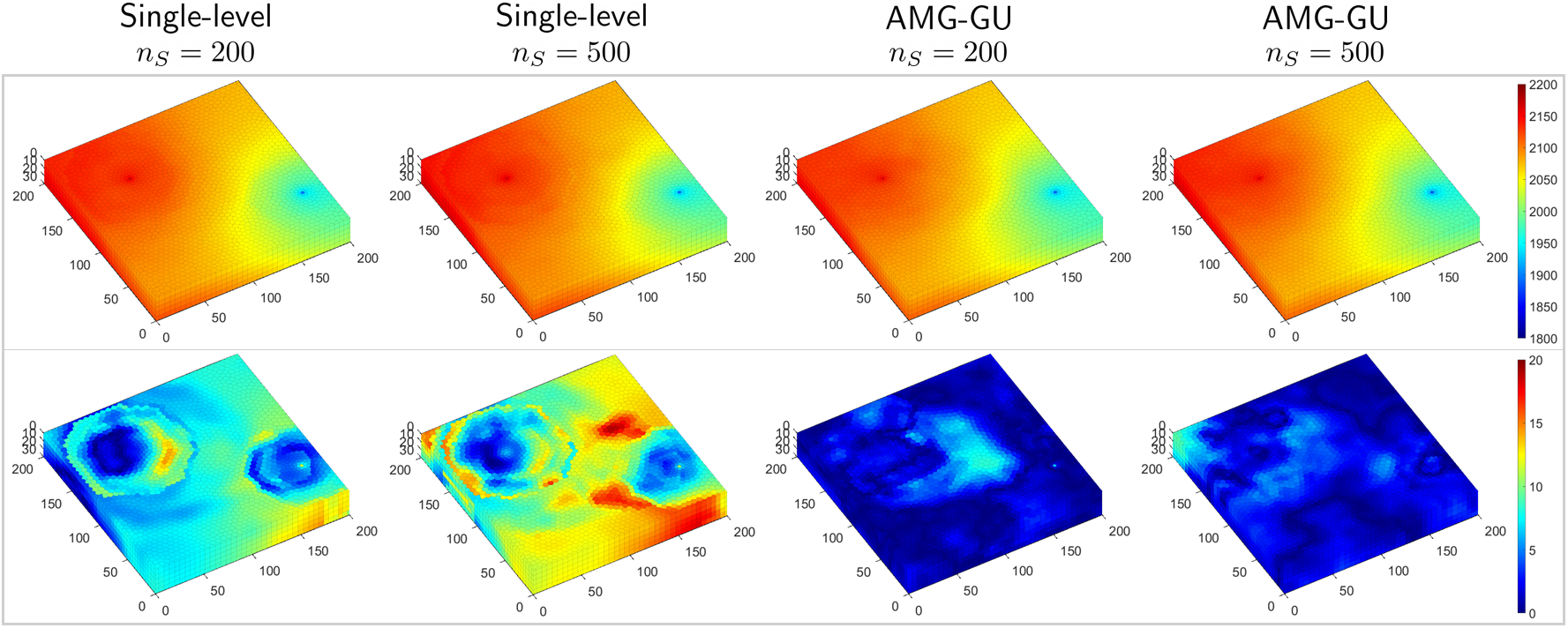}}
	\
	\subfloat[Saturation]{
		\includegraphics[scale=0.35]{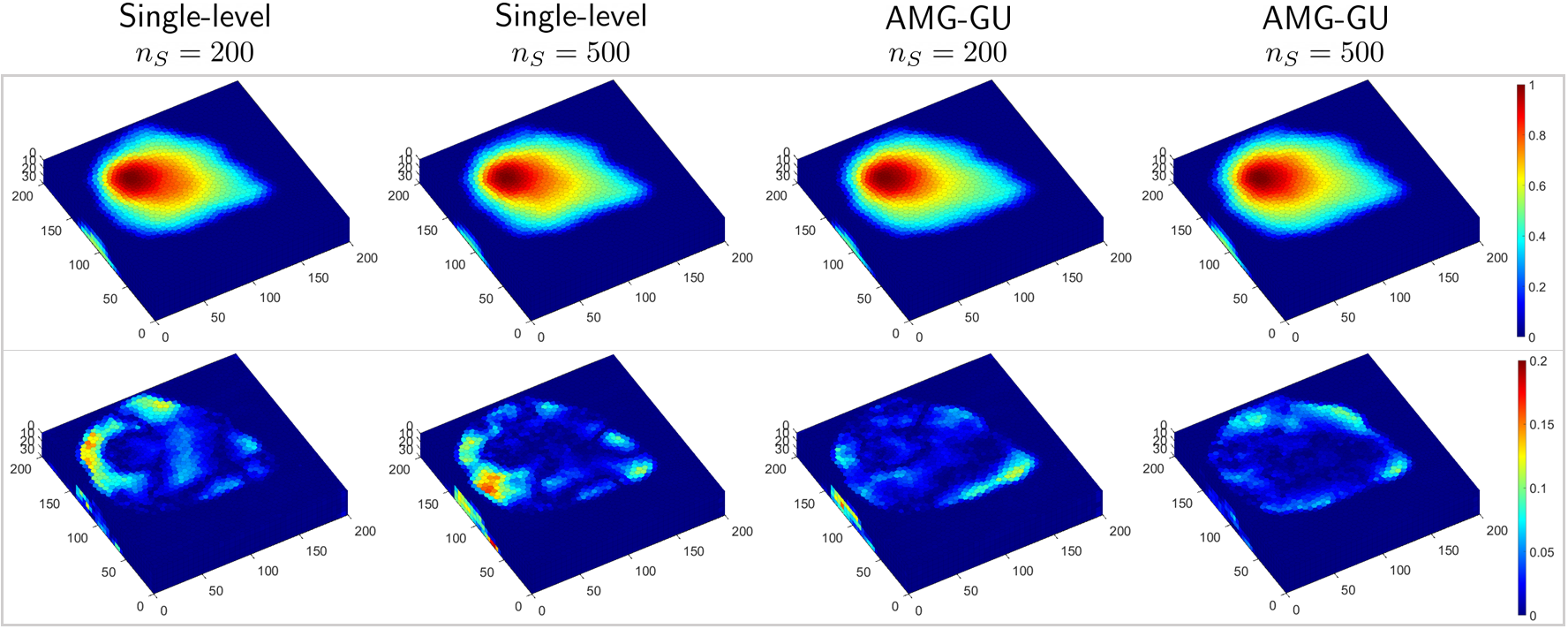}}
	\caption{Solution profiles of $\mathsf{Case\ 5}$ compared to the Single-level baseline (using 200 and 500 numbers of training samples).}
	\label{fig:sec3_5_ps}
\end{figure}

We study the influence of the training dataset size on the model performance. Qualitative comparisons between the AMG-GU model (with $\mathtt{Lloyd}$ for aggregation and knn for unpooling) and the Single-level baseline are first provided. The permeability, ground truth, and prediction fields of a representative test case are shown in \textbf{Fig.~\ref{fig:sec3_5_ps}}.

We can see that the Single-level surrogate produces unacceptable errors in the final snapshots (especially pressure) for both the scenarios using $n_S = 200$ and $n_S = 500$ numbers of training samples. In contrast, the approximations from AMG-GU excellently match (visually indistinguishable to) the ground truth. The sharp fronts as well as heterogeneous details of the fluid distribution are accurately resolved.

\begin{figure}[!htb]
	\centering
	\includegraphics[scale=0.34]{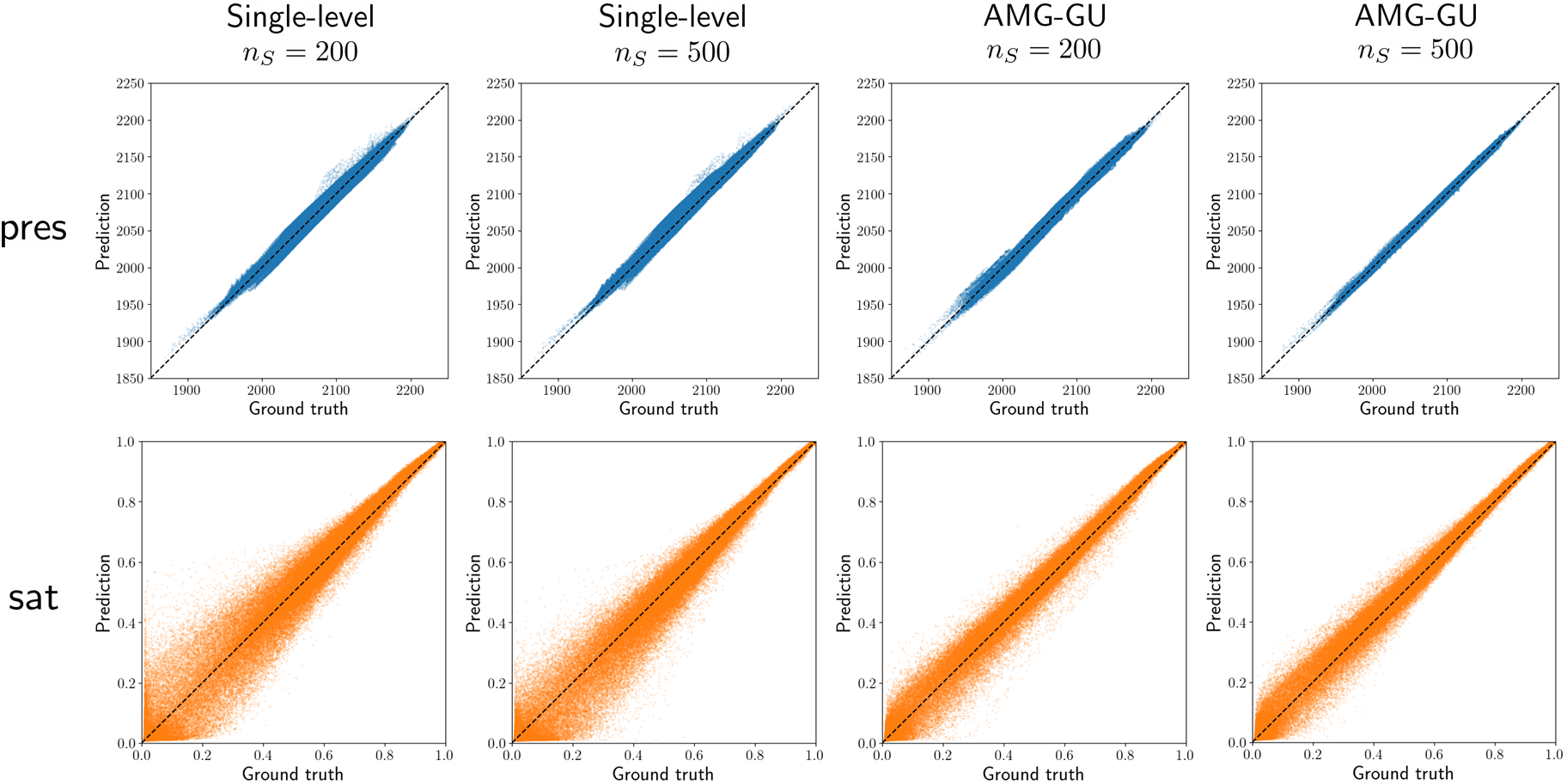}
	\caption{Parity plots of the pressure (top row) and saturation (bottom row) predictions compared to the Single-level baseline (using 200 and 500 numbers of training samples).}
	\label{fig:parity_ps}
\end{figure}

To perform quantitative evaluations, we present the parity plots over all the testing samples in \textbf{Fig.~\ref{fig:parity_ps}}, where the 45-degree line corresponds to perfect agreement. As can be seen, AMG-GU clearly outperforms the Single-level baseline, producing the points closely aligned with the ground truth across both 200 and 500 training samples.

\begin{figure}[!htb]
	\centering
	\subfloat[Pressure (psi)]{
		\includegraphics[scale=0.44]{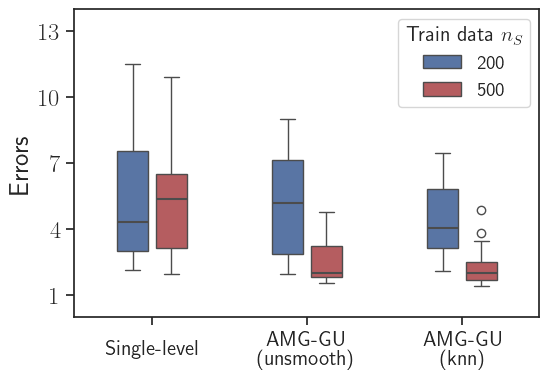}}
	\ \ 
	\subfloat[Saturation]{
		\includegraphics[scale=0.44]{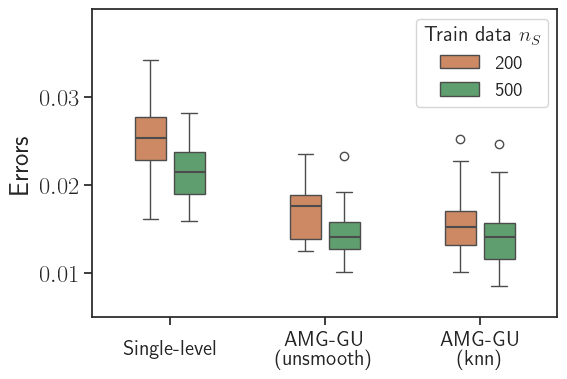}}
	\caption{Boxplots of the mean absolute pressure and saturation errors for each of the testing cases (using 200 and 500 numbers of training samples).}
	\label{fig:sec3_box_ps}
\end{figure}

We further present the boxplots of the mean absolute pressure and saturation errors for each of the testing samples in \textbf{Fig.~\ref{fig:sec3_box_ps}}. The boxplots show that AMG-GU not only achieves lower mean errors, but also narrows the spread of errors across different test cases, indicating more stable and reliable performance. 

When increasing the dataset size to 500, the two AMG-GU variants (unsmooth and knn) yield significantly smaller prediction errors. In contrast, the Single-level baseline exhibits large errors, with marginal improvement as the dataset size increases.

The primary reason for the superior performance of AMG-GU lies in its multi-level structure. The graph pooling allows for effective learning of both high-frequency local details and low-frequency global patterns, essential for the parabolic pressure dynamics. By comparison, the Single-level model operates at a fixed scale, limiting its capacity to learn long-range dependencies and thus resulting in persistent errors even with more data. The multi-scale representation provides AMG-GU with more expressiveness, enabling better generalization and more effective utilization of training data.

Overall, the above results highlight that our AMG-inspired Graph U-Net framework generalizes well to unseen well configurations, permeability fields, and unstructured meshes. The surrogate models can learn a general understanding of the complex fluid dynamics (with compressibility, gravity and capillarity) governed by the coupled multi-phase PDE system.

\section{Summary}

We rely on GNNs for surrogate modeling of multi-phase flow and transport in porous media governed by viscous, capillary, and gravity forces. Recognizing the challenges posed by the pressure component of the coupled PDEs, we developed a novel Graph U-Net model to enable hierarchical graph learning. Inspired by aggregation-type Algebraic Multigrid, we proposed a graph coarsening strategy adapted to heterogeneous PDE coefficients, achieving effective graph pooling and unpooling operations.

The performance of the trained surrogate models was evaluated using 3D heterogeneous test cases. The surrogates offer significant computational speedups—up to 160-fold—compared to the high-fidelity simulator. Both qualitative and quantitative analyses show that our multi-level surrogates predict pressure and saturation dynamics with high accuracy, significantly outperforming the baselines. The AMG-inspired Graph U-Net effectively captures both local details and long-range patterns for the pressure dynamics. The multi-level representation enhances the model generalization to unseen well configurations, permeability fields, and unstructured meshes.

\section*{Appendix A. Fully-implicit finite-volume discretization}

To solve the PDE system from Eq.~(\ref{eq:mass_con}), we apply a finite volume method that discretizes the simulation domain into a mesh consisting of $n$ cells and a fully-implicit scheme for the time discretization
\begin{equation} 
	\label{eq:dis_mass}
	\frac{\left | \Omega_i \right |}{\Delta t} \left ( \left ( \phi_i \rho_{l,i} s_{l,i} \right )^{n+1} - \left ( \phi_i \rho_{l,i} s_{l,i} \right )^{n} \right ) - \sum_{j\in adj(i)} \! \left ( \rho_{l,ij} \upsilon_{l,{ij}} \right )^{n+1} - Q_{l,i}^{n+1} = 0,
\end{equation}
where $i \in \left \{ 1,...,n \right \}$ is cell index, $\left | \Omega_i \right |$ is cell volume, $(ij)$ corresponds to the interface between cells $i$ and $j$. Superscripts represent timesteps, and $\Delta t$ is timestep size.

The discrete phase flux based on the two-point flux approximation can be written as
\begin{equation} 
	\label{eq:dis_p_f}
	\upsilon_{l,ij} = \Upsilon_{ij} \lambda_{l,ij} \Delta \Phi_{l,ij},
\end{equation}
where $\Delta \Phi_{l,ij} = \Delta p_{l,ij} - g_{l,ij}$ is the phase-potential difference with the discrete weights $g_{l,ij} = \rho_{l,ij} \, g \Delta z_{ij}$. The phase mobility $\lambda_{l,ij}$ is evaluated using the Phase-Potential Upwinding (PPU) scheme (Sammon 1988; Brenier and Jaffré 1991). In PPU, the mobility of each phase is treated separately according to the sign of the phase-potential difference. The upwinding criterion is given as
\begin{equation} 
	\label{eq:PPU}
	\lambda_{l,ij} = \left\{ {\begin{array}{*{20}c}
			\lambda_{l}(s_{i}),   & \Delta \Phi_{l,ij}\geq 0 \\ 
			\lambda_{l}(s_{j}), &  \mathrm{otherwise}
	\end{array}} \right.
\end{equation}
where $s_i = \left \{ s_{l,i} \right \}_{l \in \left \{ 1,...,n_p \right \}}$ denotes the saturations of cell $i$.

The total interface transmissibility $\Upsilon_{ij}$ combines two half-transmissibilities in a half of the harmonic average
\begin{equation}
	\Upsilon_{ij} = \frac{\Upsilon_i \Upsilon_j}{\Upsilon_i + \Upsilon_j} \, , \qquad \Upsilon_{i} = \frac{K_i A_{ij}}{d_i},
\end{equation}
where $A_{ij}$ denotes the interface area, $K_i$ is the permeability of cell $i$, and $d_i$ is the length from the cell centroid to the interface.

In the finite volume formulation, the discrete source (or sink) term for a mesh cell containing a well (referred to as well cell) is written as (Peaceman 1983)
\begin{equation} 
	Q_{l,i} = \textrm{WI}_i \, \left ( \rho_{l} \lambda_{l} \right )_i \left ( p_l - p^W \right )_i.
\end{equation}
which represents the well flux for phase $l$ in cell $i$. $p_{l,i}$ is well-cell pressure, $p^W_i$ is wellbore pressure, and $\textrm{WI}_i$ is well index. Note that $Q_{l}$ is zero everywhere except at a well cell.

\section*{Appendix B. Lloyd aggregation}

The Lloyd aggregation algorithm (\ref{alg:Lloyd_Aggregation}) is an iterative graph partitioning method (Lloyd 1982; Bell 2008) that divides the nodes \( \mathcal{X} \) of a graph into non-overlapping aggregates based on proximity to a set of center nodes \( \mathcal{X}_c \).

The algorithm begins by selecting an initial set of aggregate centers \( \mathcal{X}_c \) from \( \mathcal{X} \). Each node in the graph is then assigned to its nearest center using the modified Bellman-Ford algorithm (Leiserson et al. 1994), which computes the shortest path distances \( \bm{d} \) between nodes and their respective centers.

After the initial assignment, the algorithm proceeds to update the aggregate centers. This is done by first identifying the border nodes \( \mathcal{B} \)—nodes that lie on the boundary between different aggregates. Then modified Bellman-Ford algorithm is applied again to calculate the distances from \( \mathcal{B} \). The new center of each aggregate is selected as the node farthest from the border, ensuring that the center is well-positioned within the aggregate.

The algorithm iterates through these steps—reassigning nodes to aggregates and updating the centers—for a fixed number of iterations \( n_{it} \), or until the aggregates stabilize. The final output of the algorithm is the vector \( \bm{z} \), which indicates the final aggregate assignments for each node. The Lloyd aggregation provides a scalable and effective way for graph partitioning. Its complexity is linear in the number of nodes and edges, making it well-suited for large-scale problems.

\begin{algorithm}
	\caption{Lloyd aggregation}
	\label{alg:Lloyd_Aggregation}
	\KwIn{\\
		$n_{it}$: \ $\textrm{\small Number of iterations}$ \\ 
		$\mathcal{E}$: \: \ $\textrm{\small Set of all edges}$ \\ 
		$\mathcal{X}$: \! \! \! $\textrm{\small Set of all nodes}$ \\ 
		$\mathcal{X}_c$: \! \! $\textrm{\small Set of initial center nodes}$
	}
	\KwOut{\\
		$\bm{z}$: \, \! \! $\textrm{\small Final aggregate assignments}$
	}
	\medskip	
	\For{$\kappa = 1,2,\dots,n_{it}$ \ }{
		$\bm{d}, \bm{z} \leftarrow \mathtt{mBellmanFord}\left ( \mathcal{E}, \mathcal{X}, \mathcal{X}_c \right )$ \hfill $\{\textrm{\footnotesize compute distances and nearest centers}\}$ \\ 
		$\mathcal{B} \leftarrow \{\}$ \hfill $\{\textrm{\footnotesize initialize set of border nodes}\}$ \\ 
		\For{$(i, j) \in \mathcal{E}$}{
			\If{$z_i \neq z_j$}{
				$\mathcal{B} \leftarrow \mathcal{B} \cup \{i, j\}$ \hfill $\{\textrm{\footnotesize identify border nodes}\}$ \\ 
			}
		}
		$\bm{d}, \bm{b} \leftarrow \mathtt{mBellmanFord}\left ( \mathcal{E}, \mathcal{X}, \mathcal{B} \right )$ \hfill $\{\textrm{\footnotesize recompute distances from border nodes}\}$ \\ 
		$\mathcal{X}_c \leftarrow \{ i \in \mathcal{X}: \, d_i > d_j \ \forall \, z_i = z_j\}$ \hfill $\{\textrm{\footnotesize update center nodes}\}$ \\ 
	}
	\Return{$\bm{z}$}
\end{algorithm}

\section*{Acknowledgements}
We thank Bo Guo at The University of Arizona for constructive discussions. The work is supported by National Key R\&D Program of China (Nos.\,2022YFA1005200, 2022YFA1005202, and 2022YFA1005203), NSFC Major Research Plan -  Interpretable and General-purpose Next-generation Artificial Intelligence (No 92370205), Anhui Center for Applied Mathematics, and Key Laboratory of the Ministry of Education for Mathematical Foundations and Applications of Digital Technology, University of Science and Technology of China.

\section*{References}

Brandt, A., 1986. Algebraic multigrid theory: The symmetric case. Applied mathematics and computation, 19(1-4), pp.23-56.

Brenier, Y. and Jaffré, J., 1991. Upstream differencing for multiphase flow in reservoir simulation. SIAM journal on numerical analysis, 28(3), pp.685-696.

Bell, W.N., 2008. Algebraic Multigrid for Discrete Differential Forms. Ph.D. thesis, University of Illinois at Urbana-Champaign. 

Bahdanau, D., Cho, K. and Bengio, Y., 2014. Neural machine translation by jointly learning to align and translate. arXiv preprint arXiv:1409.0473.

Ba, J. L.; Kiros, J. R.; Hinton, G. E. Layer normalization. arXiv preprint arXiv:1607.06450, 2016.

Battaglia, P., Pascanu, R., Lai, M. and Jimenez Rezende, D., 2016. Interaction networks for learning about objects, relations and physics. Advances in neural information processing systems, 29.

Battaglia, P.W., Hamrick, J.B., Bapst, V., Sanchez-Gonzalez, A., Zambaldi, V., Malinowski, M., Tacchetti, A., Raposo, D., Santoro, A., Faulkner, R. and Gulcehre, C., 2018. Relational inductive biases, deep learning, and graph networks. arXiv preprint arXiv:1806.01261.

Bar-Sinai, Y., Hoyer, S., Hickey, J. and Brenner, M.P., 2019. Learning data-driven discretizations for partial differential equations. Proceedings of the National Academy of Sciences, 116(31), pp.15344-15349.

Bhatnagar, S., Afshar, Y., Pan, S., Duraisamy, K. and Kaushik, S., 2019. Prediction of aerodynamic flow fields using convolutional neural networks. Computational Mechanics, 64, pp.525-545.

Belbute-Peres, F.D.A., Economon, T. and Kolter, Z., 2020, November. Combining differentiable PDE solvers and graph neural networks for fluid flow prediction. In international conference on machine learning (pp. 2402-2411). PMLR.

Bianchi, F.M., Grattarola, D. and Alippi, C., 2020, November. Spectral clustering with graph neural networks for graph pooling. In International conference on machine learning (pp. 874-883). PMLR.

Bell, N., Olson, L., Schroder, J. and Southworth, B.S., 2022. PyAMG: Algebraic multigrid solvers in python. Journal of Open Source Software.

Brandstetter, J., Worrall, D. and Welling, M., 2022. Message passing neural PDE solvers. arXiv preprint arXiv:2202.03376.

Chavent, G. and Jaffré, J., 1986. Mathematical models and finite elements for reservoir simulation: single phase, multiphase and multicomponent flows through porous media. Elsevier.

Chen, Z. and Ewing, R.E., 1997. Comparison of various formulations of three-phase flow in porous media. Journal of Computational Physics, 132(2), pp.362-373.

Chen, Z., Huan, G. and Ma, Y., 2006. Computational methods for multiphase flows in porous media. Society for Industrial and Applied Mathematics.

Chen, J., Hachem, E. and Viquerat, J., 2021. Graph neural networks for laminar flow prediction around random two-dimensional shapes. Physics of Fluids, 33(12), p.123607.

Cao, Y., Chai, M., Li, M. and Jiang, C., 2022. Efficient Learning of Mesh-Based Physical Simulation with BSMS-GNN. arXiv preprint arXiv:2210.02573.

Dhillon, I.S., Guan, Y. and Kulis, B., 2007. Weighted graph cuts without eigenvectors a multilevel approach. IEEE transactions on pattern analysis and machine intelligence, 29(11), pp.1944-1957.

Deshpande, S., Bordas, S.P. and Lengiewicz, J., 2024. Magnet: A graph u-net architecture for mesh-based simulations. Engineering Applications of Artificial Intelligence, 133, p.108055.

Eliasof, M. and Treister, E., 2020. Diffgcn: Graph convolutional networks via differential operators and algebraic multigrid pooling. Advances in neural information processing systems, 33, pp.18016-18027.

Fortunato, M., Pfaff, T., Wirnsberger, P., Pritzel, A. and Battaglia, P., 2022. Multiscale meshgraphnets. arXiv preprint arXiv:2210.00612.

Franco, N.R., Fresca, S., Tombari, F. and Manzoni, A., 2023. Deep learning-based surrogate models for parametrized PDEs: Handling geometric variability through graph neural networks. Chaos: An Interdisciplinary Journal of Nonlinear Science, 33(12).

Guo, X., Li, W. and Iorio, F., 2016, August. Convolutional neural networks for steady flow approximation. In Proceedings of the 22nd ACM SIGKDD international conference on knowledge discovery and data mining (pp. 481-490).

Gilmer, J., Schoenholz, S.S., Riley, P.F., Vinyals, O. and Dahl, G.E., 2017, July. Neural message passing for quantum chemistry. In International conference on machine learning (pp. 1263-1272). PMLR.

Gao, H. and Ji, S., 2021. Graph U-Nets. IEEE Transactions on Pattern Analysis and Machine Intelligence, 44(9), pp.4948-4960.

Grattarola, D., Zambon, D., Bianchi, F.M. and Alippi, C., 2022. Understanding pooling in graph neural networks. IEEE transactions on neural networks and learning systems, 35(2), pp.2708-2718.

Ioffe, S., 2015. Batch normalization: Accelerating deep network training by reducing internal covariate shift. arXiv preprint arXiv:1502.03167.

Iakovlev, V., Heinonen, M. and Lähdesmäki, H., 2020. Learning continuous-time pdes from sparse data with graph neural networks. arXiv preprint arXiv:2006.08956.

Jiang, Z., Tahmasebi, P. and Mao, Z., 2021. Deep residual U-net convolution neural networks with autoregressive strategy for fluid flow predictions in large-scale geosystems. Advances in Water Resources, 150, p.103878.

Jiang, S. and Durlofsky, L.J., 2023. Use of multifidelity training data and transfer learning for efficient construction of subsurface flow surrogate models. Journal of Computational Physics, 474, p.111800.

Jiang, J., 2024. Simulating multiphase flow in fractured media with graph neural networks. Physics of Fluids, 36(2).

Ju, X., Hamon, F.P., Wen, G., Kanfar, R., Araya-Polo, M. and Tchelepi, H.A., 2024. Learning CO2 plume migration in faulted reservoirs with Graph Neural Networks. Computers \& Geosciences, p.105711.

Kingma, D.P. and Ba, J., 2014. Adam: A method for stochastic optimization. arXiv preprint arXiv:1412.6980.

Kipf, T.N. and Welling, M., 2016. Semi-supervised classification with graph convolutional networks. arXiv preprint arXiv:1609.02907.

Krizhevsky, A., Sutskever, I. and Hinton, G.E., 2012. Imagenet classification with deep convolutional neural networks. Advances in neural information processing systems, 25.

Lloyd, S., 1982. Least squares quantization in PCM. IEEE transactions on information theory, 28(2), pp.129-137.

Leiserson, C.E., Rivest, R.L., Cormen, T.H. and Stein, C., 1994. Introduction to algorithms (Vol. 3). Cambridge, MA, USA: MIT press.

LeCun, Y., Bottou, L., Bengio, Y. and Haffner, P., 1998. Gradient-based learning applied to document recognition. Proceedings of the IEEE, 86(11), pp.2278-2324.

Lee, J., Lee, I. and Kang, J., 2019, May. Self-attention graph pooling. In International conference on machine learning (pp. 3734-3743). pmlr.

Li, Z., Kovachki, N., Azizzadenesheli, K., Liu, B., Bhattacharya, K., Stuart, A. and Anandkumar, A., 2020. Fourier neural operator for parametric partial differential equations. arXiv preprint arXiv:2010.08895.

Li, Z., Kovachki, N., Azizzadenesheli, K., Liu, B., Bhattacharya, K., Stuart, A. and Anandkumar, A., 2020. Neural operator: Graph kernel network for partial differential equations. arXiv preprint arXiv:2003.03485.

Liu, B., Tang, J., Huang, H. and Lu, X.Y., 2020. Deep learning methods for super-resolution reconstruction of turbulent flows. Physics of fluids, 32(2).

Lu, L., Jin, P., Pang, G., Zhang, Z. and Karniadakis, G.E., 2021. Learning nonlinear operators via DeepONet based on the universal approximation theorem of operators. Nature machine intelligence, 3(3), pp.218-229.

Lino, M., Fotiadis, S., Bharath, A.A. and Cantwell, C.D., 2022. Multi-scale rotation-equivariant graph neural networks for unsteady Eulerian fluid dynamics. Physics of Fluids, 34(8).

Liu, C., Zhan, Y., Wu, J., Li, C., Du, B., Hu, W., Liu, T. and Tao, D., 2022. Graph pooling for graph neural networks: Progress, challenges, and opportunities. arXiv preprint arXiv:2204.07321.

Lam, R., Sanchez-Gonzalez, A., Willson, M., Wirnsberger, P., Fortunato, M., Alet, F., Ravuri, S., Ewalds, T., Eaton-Rosen, Z., Hu, W. and Merose, A., 2023. Learning skillful medium-range global weather forecasting. Science, 382(6677), pp.1416-1421.

Muresan, A.C. and Notay, Y., 2008. Analysis of aggregation-based multigrid. SIAM Journal on Scientific Computing, 30(2), pp.1082-1103.

Mo, S., Zhu, Y., Zabaras, N., Shi, X. and Wu, J., 2019. Deep convolutional encoder‐decoder networks for uncertainty quantification of dynamic multiphase flow in heterogeneous media. Water Resources Research, 55(1), pp.703-728.

Maldonado-Cruz, E. and Pyrcz, M.J., 2022. Fast evaluation of pressure and saturation predictions with a deep learning surrogate flow model. Journal of Petroleum Science and Engineering, 212, p.110244.

Nytko, N., 2022. Learning aggregates and interpolation for algebraic multigrid (Doctoral dissertation, University of Illinois at Urbana-Champaign).

Peaceman, D.W., 1983. Interpretation of well-block pressures in numerical reservoir simulation with nonsquare grid blocks and anisotropic permeability. Society of Petroleum Engineers Journal, 23(03), pp.531-543.

Qi, C.R., Yi, L., Su, H. and Guibas, L.J., 2017. Pointnet++: Deep hierarchical feature learning on point sets in a metric space. Advances in neural information processing systems, 30.

Paszke, A., Gross, S., Massa, F., Lerer, A., Bradbury, J., Chanan, G., Killeen, T., Lin, Z., Gimelshein, N., Antiga, L. and Desmaison, A., 2019. Pytorch: An imperative style, high-performance deep learning library. Advances in neural information processing systems, 32.

Pfaff, T., Fortunato, M., Sanchez-Gonzalez, A. and Battaglia, P.W., 2020. Learning mesh-based simulation with graph networks. arXiv preprint arXiv:2010.03409.

Peng, J.Z., Wang, Y.Z., Chen, S., Chen, Z.H., Wu, W.T. and Aubry, N., 2022. Grid adaptive reduced-order model of fluid flow based on graph convolutional neural network. Physics of Fluids, 34(8).

Ruge, J.W. and Stüben, K., 1987. Algebraic multigrid. In Multigrid methods (pp. 73-130). Society for Industrial and Applied Mathematics.

Ronneberger, O., Fischer, P. and Brox, T., 2015. U-net: Convolutional networks for biomedical image segmentation. In Medical image computing and computer-assisted intervention. pp. 234-241. Springer International Publishing.

Ribeiro, M.D., Rehman, A., Ahmed, S. and Dengel, A., 2020. DeepCFD: Efficient steady-state laminar flow approximation with deep convolutional neural networks. arXiv preprint arXiv:2004.08826.

Sammon, P.H., 1988. An analysis of upstream differencing. SPE reservoir engineering, 3(03), pp.1-053.

Scherer, D., Müller, A. and Behnke, S., 2010, September. Evaluation of pooling operations in convolutional architectures for object recognition. In International conference on artificial neural networks (pp. 92-101). Berlin, Heidelberg: Springer Berlin Heidelberg.

Simonovsky, M. and Komodakis, N., 2017. Dynamic edge-conditioned filters in convolutional neural networks on graphs. In Proceedings of the IEEE conference on computer vision and pattern recognition (pp. 3693-3702).

Sekar, V., Jiang, Q., Shu, C. and Khoo, B.C., 2019. Fast flow field prediction over airfoils using deep learning approach. Physics of Fluids, 31(5).

Sanchez-Gonzalez, A., Godwin, J., Pfaff, T., Ying, R., Leskovec, J. and Battaglia, P., 2020, November. Learning to simulate complex physics with graph networks. In International conference on machine learning (pp. 8459-8468). PMLR.

Santos, J.E., Xu, D., Jo, H., Landry, C.J., Prodanović, M. and Pyrcz, M.J., 2020. PoreFlow-Net: A 3D convolutional neural network to predict fluid flow through porous media. Advances in Water Resources, 138, p.103539.

Thuerey, N., Weißenow, K., Prantl, L. and Hu, X., 2020. Deep learning methods for Reynolds-averaged Navier–Stokes simulations of airfoil flows. AIAA Journal, 58(1), pp.25-36.

Tang, M., Liu, Y. and Durlofsky, L.J., 2020. A deep-learning-based surrogate model for data assimilation in dynamic subsurface flow problems. Journal of Computational Physics, 413, p.109456.

Vanek, P., Mandel, J. and Brezina, M., 1996. Algebraic multigrid by smoothed aggregation for second and fourth order elliptic problems. Computing, 56(3), pp.179-196.

Veličković, P., Cucurull, G., Casanova, A., Romero, A., Lio, P. and Bengio, Y., 2017. Graph attention networks. arXiv preprint arXiv:1710.10903.

Wen, G., Tang, M. and Benson, S.M., 2021. Towards a predictor for CO2 plume migration using deep neural networks. International Journal of Greenhouse Gas Control, 105, p.103223.

Wen, G., Li, Z., Azizzadenesheli, K., Anandkumar, A. and Benson, S.M., 2022. U-FNO—An enhanced Fourier neural operator-based deep-learning model for multiphase flow. Advances in Water Resources, 163, p.104180.

Wang, L., Fournier, Y., Wald, J.F. and Mesri, Y., 2023. A graph neural network-based framework to identify flow phenomena on unstructured meshes. Physics of Fluids, 35(7).

Yan, D.M., Wang, W., Lévy, B. and Liu, Y., 2013. Efficient computation of clipped Voronoi diagram for mesh generation. Computer-Aided Design, 45(4), pp.843-852.

Ying, Z., You, J., Morris, C., Ren, X., Hamilton, W. and Leskovec, J., 2018. Hierarchical graph representation learning with differentiable pooling. Advances in neural information processing systems, 31.

Yan, B., Harp, D.R., Chen, B. and Pawar, R., 2022. A physics-constrained deep learning model for simulating multiphase flow in 3D heterogeneous porous media. Fuel, 313, p.122693.

Zhang, M., Cui, Z., Neumann, M. and Chen, Y., 2018, April. An end-to-end deep learning architecture for graph classification. In Proceedings of the AAAI conference on artificial intelligence (Vol. 32, No. 1).

Zhang, K., Zuo, Y., Zhao, H., Ma, X., Gu, J., Wang, J., Yang, Y., Yao, C. and Yao, J., 2022. Fourier neural operator for solving subsurface oil/water two-phase flow partial differential equation. Spe Journal, 27(03), pp.1815-1830.

\end{document}